\documentclass[journal]{IEEEtran} %%双栏
\IEEEoverridecommandlockouts %showing the thanks at the left bottom of the first page
\usepackage{epsfig,latexsym}
\usepackage{float}
\usepackage{indentfirst}
\usepackage{graphicx}
\usepackage{float}
\usepackage{amsmath}
\usepackage{amssymb}
\usepackage{times}
\usepackage{subfigure}
\usepackage{algorithm}
\usepackage{algorithmic}
\usepackage{pifont}
\usepackage{psfrag}
\usepackage{cite}
\usepackage{url}
\usepackage{lastpage}

\usepackage{bbding}
\usepackage{amsmath}
\usepackage{amssymb}
\usepackage{ntheorem}
\usepackage{graphicx}
\usepackage{xcolor}
\psfull

\begin{document}

\title{Robust Designs of Beamforming and Power\\
 Splitting for Distributed Antenna Systems\\
with Wireless Energy Harvesting}

\author{Zhengyu Zhu, Sai Huang, Zheng Chu, Fuhui Zhou,\\
 Di Zhang and Inkyu Lee, \emph{Fellow, IEEE}
%\thanks{This work was supported  by the National
%Nature Science Foundation of China under grant (61571402, 61601516, 61701214), the Young Natural Science Foundation of Jiangxi Province (20171BAB212002), the China Postdoctoral Science Foundation (2017M610400), the Postdoctoral Science Foundation of Jiangxi Province(2017KY04), and by the National Research Foundation through the Ministry of Science, ICT, and Future Planning (MSIP), Korean Government under Grant 2017R1A2B3012316.}
\thanks{Z. Zhu is with the School of Information Engineering,
Zhengzhou University, Zhengzhou 450001, China (e-mail: zhuzhengyu6@gmail.com). S. Huang is with the School
of Information and Communication Engineering, Beijing University of posts and telecommunications, Beijing, China (e-mail: huangsai@bupt.edu.cn). Z. Chu is  with the 5G Innovation Center (5GIC), Institute of Communication Systems (ICS), University of Surrey, Guildford, GU2 7XH, United Kingdom. (email: andrew.chuzheng7@gmail.com).  F. Zhou is with Information Engineering School, Nanchang University,  Nanchang, China. (email: zhoufuhui1989@163.com). {D. Zhang is with the School of Information Engineering, Zhengzhou University, Zhengzhou, 450-001, China, and also with the Information System Laboratory, Department of Electrical and Computer Engineering, Seoul National University, Seoul, 151-744, Korea (e-mail: di\_zhang@islab.snu.ac.kr).} I. Lee is with School of Electrical Engineering, Korea University, Seoul, Korea (e-mail: inkyu@korea.ac.kr).}}

\maketitle
%%%%%%%%%%%%%%%%%%%%%%%%%%%%%%%%%%%%%%%%%%%%%%%%%%%%%%%%%%%%%
\begin{abstract}
%%%%%%%%%%%%%%%%%%%%%%%%%%%%%%%%%%%%%%%%%%%%%%%%%%%%%%%%%%%%%
In this paper, we investigate a multiuser  distributed antenna system with simultaneous wireless information and power transmission under the assumption of imperfect channel state information (CSI). In this system, a distributed antenna port with multiple antennas supports a set of mobile stations who can decode information and harvest energy simultaneously via a power splitter. To design robust transmit beamforming vectors and the power splitting (PS) factors in the presence of CSI errors, we maximize the average worst-case signal-to-interference-plus-noise ratio (SINR) while achieving individual energy harvesting constraint for each mobile station.  First, we develop an efficient algorithm to convert the max-min SINR problem to a set of ``dual" min-max power balancing problems. Then, motivated by the penalty function method, an iterative algorithm based on semi-definite programming (SDP) is proposed to achieve a local optimal rank-one solution. Also, to reduce the computational complexity, we present another iterative scheme based on the Lagrangian method and the successive convex approximation (SCA) technique to yield a suboptimal solution. Simulation results are shown to validate the robustness and effectiveness of the proposed algorithms.
\end{abstract}

%%%%%%%%%%%%%%%%%%%%%%%%%%%%%%%%%%%%%%%%%%%%%%%%%%%%%%%%%%%%%
\section{Introduction}
%%%%%%%%%%%%%%%%%%%%%%%%%%%%%%%%%%%%%%%%%%%%%%%%%%%%%%%%%%%%%
For the past decade, there has been a considerable evolution of wireless networks to satisfy demands on high speed data. Since resources shared among users are limited, a capacity increase is technically challenging in the wireless networks. Recently, a distributed antenna system (DAS) has received a lot of attentions as a new cellular communication structure to expand coverage and increase sum rates {\cite{jingdian1_DAS,jingdian2_DAS,jingdian3_DAS}.}

Unlike conventional cellular systems where all antennas are co-located at the cell center, distributed antenna (DA) ports of the DAS are separated geographically in a cell and are connected with each other by backhaul links \cite{Lee_12TWC_DAS}. Each DA port in the DAS is usually equipped with its own power amplifier at the analog front-end \cite{Lee_12TWC_DAS} \cite{Lee_13TWC_DAS}. Thus, individual power constraint at each antenna should be considered for the DAS unlike the conventional systems which normally impose sum power constraint \cite{Lee_13TWC_DAS}.

In the meantime, one of the limits in current cellular communication systems is the short lifetime of batteries.
To combat the battery problem of mobile users, simultaneous wireless information and power
transmission (SWIPT) has been studied in {\cite{SWIPT_ISIT2008,SWIPT_ISIT2010,RZhang_13TWC_SWIPT_PS,zhu_16JCN_SWIPT_PS,NG_14TWC_SWIPT_PS,Shi_14TWC_SWIPT_PS,zhu_16ICC_SWIPT_PS,TIFS_zzy}}. With the aid of the SWIPT,  users can charge their devices based on the received signal \cite{RZhang_13TWC_SWIPT_PS} \cite{zhu_16JCN_SWIPT_PS}.
To realize the SWIPT, a co-located receiver has been proposed  \cite{NG_14TWC_SWIPT_PS}, which employs a power splitter to perform energy harvesting (EH) and information decoding (ID) at the same time \cite{Shi_14TWC_SWIPT_PS}. By adopting the power splitting (PS) receiver, the SWIPT scheme for multiple-input single-output (MISO) downlink systems has been examined in \cite{RZhang_13TWC_SWIPT_PS} and \cite{Shi_14TWC_SWIPT_PS} where perfect channel state information at the transmitter (CSIT) was assumed. In practice, however, due to channel estimation errors and feedback delays, it is not possible to obtain perfect CSIT \cite{Chu_16TWC_Secrecy,Chu_16TVT_SWIPT,zhu_IET_robust_SWIPT,CL_Chu}.

On the other hand, some recent works have investigated SWIPT in DAS \cite{Jin_SWIPT_DAS_15CM,Ding_SWIPT_smart_15CM,Jin_SWIPT_DAS_16arxiv, HuangKB_SWIPTDAS_16GLOBECOM,NG_16TWC_SWIPT_DAS,zhu_15VTC_SWIPT_DAS,Magazine_Huangkaibin,ICNC_DAS}.
\cite{Jin_SWIPT_DAS_15CM} has provided several intuitions and revealed
the challenges and opportunities in DAS SWIPT systems. In order to improve energy efficiency of SWIPT, the application of advanced smart antenna technologies has been focused in \cite{Ding_SWIPT_smart_15CM}.
In \cite{Jin_SWIPT_DAS_16arxiv},  a power management strategy has been studied to supply maximum wireless information transfer (WIT) with
minimum wireless energy transfer (WET) constraint for adopting PS. Moreover, a tradeoff between
the power transfer efficiency and the information transfer capacity has been introduced in \cite{HuangKB_SWIPTDAS_16GLOBECOM}.
The work in \cite{zhu_15VTC_SWIPT_DAS} examined a design of robust beamforming and PS for multiuser downlink DAS SWIPT. However, only one antenna was considered in each DA port. The authors in \cite{NG_16TWC_SWIPT_DAS} investigated resource allocation for DAS SWIPT systems  based on the worst-case model, where per-DA port power constraint was adopted.
In \cite{Magazine_Huangkaibin}, a few open issues and promising research trends in the wireless powered communications area with DAS were introduced.
In addition, to achieve a balance between transmission power and circuit power, \cite{ICNC_DAS}  studies a system utility minimization problem in a DAS SWIPT system via joint design of remote radio heads selection and beamforming.
However, joint optimal design of transmit beamforming and the receive PS factor for SWIPT
in DAS PS-based systems with multiple transmit antennas of each DA port, has not been considered in the literature yet.

{Motivated by the existing literature \cite{Jin_SWIPT_DAS_15CM,Ding_SWIPT_smart_15CM,Jin_SWIPT_DAS_16arxiv, HuangKB_SWIPTDAS_16GLOBECOM,NG_16TWC_SWIPT_DAS,zhu_15VTC_SWIPT_DAS,Magazine_Huangkaibin,ICNC_DAS},} in this paper, we study a joint design of robust transmit beamforming at the DA port and the receive PS factors at mobile stations (MSs) in multiuser DAS SWIPT systems with imperfect CSI. Channel uncertainties are modeled by the worst-case model as in \cite{zhu_15VTC_SWIPT_DAS}.
Our aim is to maximize the worst-case signal-to-interference-and-noise ratio (SINR) subject to EH constraint and per-DA port power constraint.  {The contributions of this work are summarized
as follows:}

\begin{itemize}
\item {For a given SINR target, the original problem is decomposed into a sequence of min-max per-DA port power balancing problems. In order to convert the non-convex constraint into linear matrix inequality (LMI), Schur complement is used to derive the equivalent forms of the SINR constraint and the EH constraint. Furthermore, we prove that a solution of the relaxed semi-definite program (SDP) is always rank-two. Also, to recover a near-optimal rank-one solution, we employ a penalty function method  instead of the conventional Gaussian randomization (GR) technique.}

\item    {To reduce the computational complexity,  another formulation is expressed for the minimum SINR maximization problem. By employing the
Lagrangian multiplier method and the first order Taylor expansion, the SINR constraint can be approximately reformulated into two convex forms with linear constraints.
Then, we propose an iterative algorithm based on the successive convex approximation (SCA) to find a suboptimal solution.}
\end{itemize}

{Simulations evaluation have been conducted to provide
the robustness and effectiveness of the proposed algorithms. The performance
is also compared with other recent conventional
schemes in this area. We show that the proposed algorithms
has the superior performances in terms of average worst-case rate by extensive simulation results.}

The remainder of this paper is organized as follows: in Section II, we describe a system model for the multiuser DAS SWIPT and
formulate the worst-case SINR maximization problem subject to per-DA port power and EH constraint. Section III derives the proposed robust joint designs. In Section IV, we present the computational complexity of the proposed algorithms. Simulation results are presented in Section V. Finally, Section VI concludes this paper.

\emph{\textbf{Notation:}} Lower-case letters are denoted by scalars, bold-face lower-case letters are used for vectors, and boldface
upper-case letters means matrices.  $ \|{\textbf{x}}\|$ represents the Euclidean norm of a complex vector $\textbf{x}$ and ${\rm{diag}}(\textbf{\emph{x}})$ denotes the diagonal matrix whose diagonal element vector is $\textbf{\emph{x}}$. $|z|$ stands for the norm of a complex number $z$. For a matrix $\textbf{M}$, $\textbf{M}^T$, $\textbf{M}^H$, ${\rm{rank}}(\textbf{M})$, and $[\textbf{M}]_{i,j}$ are defined as trace, transpose, conjugate transpose, rank, and the $(i,j)$-th element, respectively. $\lambda_{max}(\textbf{{M}})$ denotes the maximum eigenvalue of $\textbf{{M}}$, and $\emph{vec}(\textbf{M})$ stacks the elements of $\textbf{M}$ in a column vector. $\textbf{I}$ defines an identity matrix. $\mathbb{C}^{M\times N}$, $ \mathbb{H}^{M\times N}$ and $ \mathbb{R}^{M\times N}$ are the set of complex matrices, Hermitian matrices and real matrices of size $M\times N$,  respectively. $ \mathbb{H}_{+}$ equals the set of positive semi-definite (PSD) Hermitian matrices. $\mathbf{0}_{M\times L}$ is a null matrix with size ${M\times L}$.

%%%%%%%%%%%%%%%%%%%%%%%%%%%%%%%%%%%%%%%%%%%%%%%%%%%%%%%%%%%%%
\section{System Model and Problem Formulation}
%%%%%%%%%%%%%%%%%%%%%%%%%%%%%%%%%%%%%%%%%%%%%%%%%%%%%%%%%%%%%
In Fig. 1, we describe a single cell system model for the multiuser downlink DAS scenario with SWIPT. The DAS consists of $M$ DA ports and $K$ single-antenna MSs.  %denoted by MS$_1,...,$ MS$_K$.
It is assumed that each DA port is equipped with $N_T$ antennas, which have individual power constraint.  All DA ports are physically connected to the main processing unit {(MPU)} through fiber optics or an exclusive radio frequency (RF) link. Furthermore, all DA ports share the information of user distance and user data, but do not require CSI of all MSs as in \cite{Lee_12TWC_DAS}. The MS distance information can be simply obtained by measuring the received signal strength indicator \cite{Lee_13TWC_DAS}. Note that one MS can be supported by several DA ports.

We consider the channel model for DAS which contains both small scale and large scale fading \cite{Lee_13TWC_DAS}. We denote the channel between the $m$-th DA port $(m=1,...,M)$ and the $k$-th MS $(k=1,...,K)$  as ${{\textbf{h}}_{m,k}} = d_{m,k}^{ - {\gamma  \mathord{\left/ {\vphantom {\alpha  2}} \right. \kern-\nulldelimiterspace} 2}}{\bar {\textbf{{h}}}_{m,k}}$, where ${d_{m,k}}$ stands for the distance between the $m$-th DA port and the $k$-th MS, $\gamma$ indicates the path loss exponent, and ${\bar{\textbf{{h}}}_{m,k} \in \mathbb{C}^{N_T \times 1}}$ equals the channel vector for small scale fading.
%The elements of ${\bar{\textbf{{h}}}_{m,k} }$ are independent and identically distributed (i.i.d.) complex Gaussian random variables with zero mean and unit variance.
 For the $k$-th MS, the channel vector is given as ${{{{\textbf{h}}}}_k} = [{{ {{{\textbf{h}}}}}_{1,k}^T},..., {{ {{{\textbf{h}}}}}_{M,k}^T}]^T$.

Due to channel estimation and quantization errors, CSI is imperfect at each DA port and we assume that the
uncertainty of the channel vectors is determined by ${{\cal H}_{k}}$ as an Euclidean ball  \cite{NG_14TWC_SWIPT_PS} \cite{Chu_16TWC_Secrecy} as
\begin{equation} \label{p1}
{{\cal H}_{k}} {\rm{=}} \left\{ {\hat{{\textbf{h}}}_{k} + \Delta {\textbf{h}}_{k} \left| { {\Delta {\textbf{h}}_{k}^H }{\bf{\Phi}}_{k} {\Delta {\textbf{h}}_{k} } } \right. \le {\varepsilon_{k}^2}} \right\}, k = 1,2,...,K
\end{equation}
where the ball is centered around the actual value of the estimated CSI vector ${\hat{{\textbf{h}}}_{k}}$ from $M$ DA ports to the $k$-th MS,  $\Delta {\textbf{h}}_{k} \in {\mathbb{C}^{MN_T\times 1}}$ is the norm-bounded uncertainty vector, ${\bf{\Phi}}_{k} \in \mathbb{C}^{MN_T\times MN_T}$ defines the orientation of the region, and ${\varepsilon _{k}}$ represents the radius of the ball. % as $\Delta{{ {{\textbf{h}}}}_k} = [\Delta{{{h}}_{1,k}},..., \Delta{{ {h}}_{M,k}}]^T$. In this model, the channel is given as $h_{m,k} = \hat{h}_{m,k} + \Delta h_{m,k} \in {{\cal H}_k}$ .

\begin{figure}[!t]
\begin{center}
\includegraphics[width=3.5in]{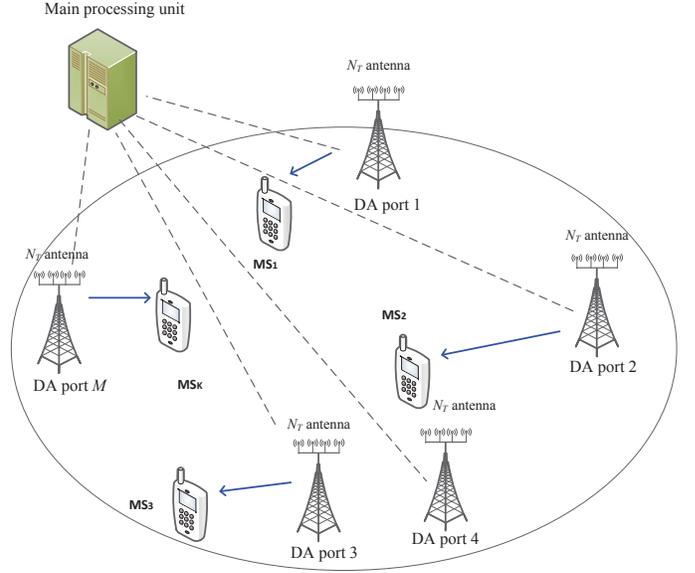}
\end{center}
\caption{Structure of a multi-user DAS downlink system}
\end{figure}

During one time slot, $K$ independent signal streams are conveyed simultaneously to $K$ MSs.  Specifically, the transmit beamforming vector ${{\textbf{v}}_{k}^m} \in {\mathbb{C}^{N_T\times 1}}$ is allocated for the $k$-th MS at the $m$-th DA port. Thus, we denote the joint  transmit beamformer vector ${{\textbf{v}}_{k}} \in {\mathbb{C}^{MN_T\times 1}}$ used by $M$ DS ports for the $k$-th MS as
$ {\textbf{{v}}_k} = \emph{vec}\left( {\left[ {\begin{array}{*{20}{c}} {{\textbf{{v}}_{1,k}}}&{{\textbf{{v}}_{2,k}}}& \ldots &{{\textbf{{v}}_{M,k}}} \end{array}} \right]} \right)$. Then, the transmitted signal to the $k$-th MS is obtained by
\begin{equation*} \label{p2}
\textbf{x}_k =  {{{\textbf{v}}_{k}}{s_{k}}}, ~\forall k,
\end{equation*}
where ${s_k} \sim {\cal C}{\cal N}\left( {0,1} \right)$ indicates the corresponding transmitted data symbol for the $k$-th MS, which is independent and identically
distributed (i.i.d.) circularly symmetric complex Gaussian (CSCG) random variable with zero mean and unit variance. We assume that each DA port has its own power constraint ${P_m} {\kern 2pt} (m = 1,...,M)$.
Let us define an $MN_T \times MN_T$ square matrix ${{{\textbf{D}}}_m} \triangleq {\rm{diag}}(\underbrace {0,...,0}_{(m - 1){N_T}},\underbrace {1,...,1}_{{N_T}},\underbrace {0,...,0}_{(M - m){N_T}})$.  Then, per-DA per power constraint is given as $\sum\limits_{k = 1}^K {{\rm{tr}}(\textbf{D}_m{{\textbf{{v}}_k}\textbf{{v}}_k^H}})  \le \alpha {P_m}, \forall m$.

The received signal at the $k$-th MS is expressed as
\begin{equation*} \label{p3}
{y_k} = {\textbf{h}}_k^H\sum\limits_{j = 1}^K {{{\textbf{v}}_j}{s_j}}  + {{\emph{n}}_k},
\end{equation*}
where ${{\emph{n}}_k} $ represents the additive white gaussian noise (AWGN) with variance $\sigma _k^2$ at the $k$-th MS. It is also assumed that each MS splits the received signal power into two parts using a power splitter, one for the EH and the other for the ID \cite{RZhang_13TWC_SWIPT_PS} \cite{Shi_14TWC_SWIPT_PS}. The PS divides the ${\rho _k} \in (0,1]$ portion and the $1- \rho_k$ portion of the received signal power to the ID and the EH, respectively.

Therefore, the split signal for the ID of the $k$-th MS is written as
\begin{equation*} \label{p4}
y_k^{ID} = \sqrt {{\rho _k}} \big( {{\textbf{h}}_k^H\sum\limits_{j = 1}^K
{{{\textbf{v}}_j}{s_j}}  + {\emph{{n}}_k}} \big) + {\emph{{z}}_k},
\end{equation*}
where ${\emph{{z}}_k} $ stands for the AWGN with variance $\delta _k^2$ during the ID process at the $k$-th MS. Then, the received SINR for the $k$-th MS is defined as
\begin{equation} \label{p5}
{\rm{SIN}}{{\rm{R}}_k}(\{\textbf{{v}}_k\},\rho_k) =  \frac{{{\rho _k}{{| {{\textbf{h}}_k^H{{\textbf{v}}_k}}|}^2}}}{{{\rho _k}\sum\limits_{j \ne k} {{{| {{\textbf{h}}_k^H{{\textbf{v}}_j}} |}^2}}  + {\rho _k}\sigma _k^2 + \delta _k^2}}.
\end{equation}

Also, due to the broadcast nature of wireless channels, the energy carried by all signals, i.e., the $1 - {\rho _k}$ portion of ${{\textbf{v}}_k}$, can be harvested at the $k$-th MS, and the split signal for the EH of the $k$-th MS is thus given as
\begin{equation*} \label{p6}
y_k^{EH} = \sqrt {1 - {\rho _k}} \big( {{\textbf{h}}_k^H\sum\limits_{j =
1}^K {{{\textbf{v}}_j}{s_j}}  + {{\emph{n}}_k}} \big).
\end{equation*}
Then, the harvested energy by the EH of the $k$-th MS is obtained as %proportional to the split signal as
\begin{equation*} \label{p7}
\displaystyle {E_k} = {\zeta _k}\left( {1 - {\rho _k}} \right)\big(
{\sum\limits_{j = 1}^K {{{| {{\textbf{h}}_k^H{{\textbf{v}}_j}} |}^2} + \sigma
_k^2} } \big)
\end{equation*}
where ${\zeta _k} \in \left( 0 \right.{\kern 1pt}\left. 1 \right]$ is the constant that
accounts for the energy conversion efficiency for the EH of the $k$-th MS.

In this paper, we assume that the harvested power at each MS should be larger than a given threshold, and each DA port also needs to satisfy per-DA port power constraint. Hence, our aim is to jointly optimize the transmit beamforming vector and the PS factor by maximizing the minimum SINR subject to EH constraint and per-DA power constraint. Then, by incorporating the norm-bounded channel uncertainty model in \eqref{p1}, the robust optimization problem is
expressed as
\begin{subequations} \label{p8}
\begin{eqnarray}
&& \!\!\!\!\!\!\!\!\!\!\!\!\!\!  \mathop {\max }\limits_{{\{{{\textbf{v}}}_{k}\}, { {\kern 1pt} {{\rho _k}} }}} {\kern 1pt} \mathop {\min }\limits_{{\textbf{h}_k} \in {{\cal H}_k}}   {\kern 7pt} {\rm{SIN}}{{\rm{R}}_k}(\{\textbf{{v}}_k\},\rho_k) \label{p8a}\\
&& \!\!\!\!\!\!\!\!\!\!\!\! {\rm{s.t.}}{\kern 6pt} {\kern 1pt} {\kern 1pt} {\kern 1pt} {\kern 1pt} {\kern 1pt} {\zeta _k}\left( {1 - {\rho _k}} \right)( { \sum\nolimits_{j = 1}^K {{{| {{{\textbf{h}}}_{k}^H{{{\textbf{v}}}_{j}}} |}^2} + \sigma _k^2} } ) \ge {e_k}, ~\forall k, \label{p8b}\\
&& {\kern 5pt} \sum\nolimits_{k = 1}^K {{\rm{tr}}(\textbf{D}_m{{\textbf{{v}}_k}\textbf{{v}}_k^H}})  \le  {P_m}, ~\forall m,  \label{p8c} \\
&& {\kern 5pt} 0 < {\rho _k} \leq 1, ~\forall k, \label{p8d}
\end{eqnarray}
\end{subequations}
%%% shuanglan
%\begin{subequations} \label{p8}
%\begin{eqnarray}
%&& \mathop {\max }\limits_{{\{{{\textbf{v}}}_{k}\}, { {\kern 1pt} {{\rho _k}} }}} {\kern 1pt} \mathop {\min }\limits_{{\textbf{h}_k} \in {{\cal H}_k}}   {\kern 7pt} {\rm{SIN}}{{\rm{R}}_k}(\{\textbf{{v}}_k\},\rho_k) \label{p8a}\\
%&&\!\!\!\!\!\! {\rm{s.t.}}{\kern 1pt} {\kern 1pt} {\kern 1pt} {\kern 1pt} {\kern 1pt} {\kern 1pt} {\zeta _k}\left( {1 - {\rho _k}} \right)( { \sum\nolimits_{j = 1}^K {{{| {{{\textbf{h}}}_{k}^H{{{\textbf{v}}}_{j}}} |}^2} + \sigma _k^2} } ) \ge {e_k}, \forall k, \label{p8b}\\
%&&\!\!\!\!\!\! {\kern 10pt} \sum\nolimits_{k = 1}^K {{\rm{tr}}(\textbf{D}_m{{\textbf{{v}}_k}\textbf{{v}}_k^H}})  \le  {P_m}, \forall m,  \label{p8c} \\
%&&\!\!\!\!\!\! {\kern 10pt} 0 < {\rho _k} \leq 1, ~\forall k. \label{p8d}
%\end{eqnarray}
%\end{subequations}
where  ${e_k}$ represents the required harvested power of the $k$-th MS.
Problem \eqref{p8} is non-convex due to  coupled variables $\left\{ {{\rho _k}} \right\}$ and $\{{{\textbf{v}}}_{k}\}$ in both the objective function and the EH constraint, and thus, is difficult to solve efficiently.

%%%%%%%%%%%%%%%%%%%%%%%%%%%%%%%%%%%%%%%%%%%%%%%%%%%%%%%%%%%%%
\section{Proposed Robust Joint Designs}
In this section, we propose two robust joint design algorithms for problem \eqref{p8}. First, we present a bisection search method which generates a local optimal rank-one solution. To reduce the computational complexity, we then introduce an SCA based algorithm to achieve a suboptimal solution.

\subsection{Proposed  Method Based on Bisection Search}
%%%%%%%%%%%%%%%%%%%%%%%%%%%%%%%%%%%%%%%%%%%%%%%%%%%%%%%%%%%%%
To make problem \eqref{p8} tractable, we decompose the problem into a set of the min-max per-DA port power balancing problems, one for each given SINR target $\Gamma > 0$ \cite{Chu_16TVT_SWIPT}.  Using bisection search over $\Gamma$, the optimal solution to problem \eqref{p8} can be obtained by solving the corresponding min-max per-DA port power balancing problem with different $\Gamma$. Then, for a given $\Gamma$, we focus on the following min-max per-DA port power balancing problem as
\begin{subequations} \label{p9}
\begin{eqnarray}
&&\!\!\!\!\!\!\!\! \displaystyle \mathop {\min }\limits_{\{{{\textbf{v}}}_{k}\},\rho_k} ~~\mathop {\max }\limits_{1 \le m \le M}  {\kern 6pt} \frac{\sum\nolimits_{k = 1}^K {{\rm{tr}}(\textbf{D}_m{{\textbf{{v}}_k}\textbf{{v}}_k^H}})}{{{P_m}}} \label{p9a}\\
&&\!\!\!\!\!\!\!\! \displaystyle  {\kern 1pt}  {\rm{s.t.}} {\kern 7pt}  {\zeta _k}\left( {1 - {{ \rho }_k}} \right)( {\sum\nolimits_{j = 1}^K {{{| {{{\textbf{h}}}_{k}^H{{{\textbf{v}}}_{j}}} |}^2} + \sigma _k^2} } ) \ge {e_k}, ~\forall k, \label{p9b}\\
&& {\kern 5pt} \displaystyle   { {\rm{SIN}}{{\rm{R}}_k}}(\{\textbf{{v}}_k\},\rho_k) \ge \Gamma, ~\forall k,  \label{p9c} \\
&& {\kern 5pt} 0 < {\rho _k} \leq 1, ~\forall k. \label{p9d}
\end{eqnarray}
\end{subequations}

We represent ${\alpha ^*}\left( \Gamma \right)$ as the optimal objective value of problem \eqref{p9}.
%We define ${{{\textbf{W}}}^{\rm{*}}}\left( \Gamma  \right)= [\textbf{\emph{w}}_1^*\left( \Gamma  \right),...,\textbf{\emph{w}}_K^*\left( \Gamma \right)]$ as the
%optimal robust beamforming matrix for problem (10) with a given $\Gamma$.
%It is proved that the optimal value ${\alpha ^*}\left( \Gamma\right)$ is a strictly monotonicity increasing function of $\Gamma$. %The following lemma explains the property of ${\alpha ^*}\left( \lambda  \right)$.
%%For this function, we have the following lemma.
% The following lemma further reveal the relationship between problems (7) and (10).
%
%{\emph{{\underline{\textbf{Lemma 1}}}:}} Denoting $\lambda_o$ as a solution of ${\alpha ^*(\lambda_o)}=1$ in problem (10), $\lambda_o$ and  ${{{\textbf{W}}}^{\rm{*}}} (\lambda_o)$ are the optimal value and the robust beamforming solution for problem (7), respectively.
%
%%Denoting $\lambda_o$ as a solution of ${\alpha ^*(\lambda_o)}=1$ and the  the corresponding ${{\bf{V}}^{\rm{*}}} (\lambda_o)$ are the optimal value and robust beamforming solution for problem (8), respectively.
%
%\emph{{\underline{Proof}:}} See Appendix A.
Note that based on the equation ${\alpha ^{\rm{*}}}( \Gamma )=1$ \cite[Lemma 2]{zhu_15VTC_SWIPT_DAS},  we can obtain the optimal beamforming solution for problem \eqref{p8}.
%To solve this equation, we first need to seek ${\alpha
%^{\rm{*}}}( \lambda_o  )$ for problem (10) with a given $\lambda $.
%Obviously, the optimal objective value of problem \eqref{p9} is monotonically non-increasing with respect to the optimal value $\Gamma^*$.
{Problem \eqref{p9}  is still non-convex in terms of the non-convex objective function \eqref{p9a}. First, we tackle the objective function \eqref{p9a} by introducing an auxiliary variable $\alpha$.} Then, the min-max per-DA port power balancing problem \eqref{p9} can be rewritten as
\begin{subequations} \label{p10}
\begin{eqnarray}
&& \mathop {\min }\limits_{\{{{\textbf{v}}}_{k}\},{\kern 1pt}{{ \rho }_k},{\kern 1pt}\alpha ,{\kern 1pt}{\textbf{h}}_k \in {{\cal H}_k}} {\kern 31pt}  \alpha  \label{p10a}\\
&& {\rm{s}}{\rm{.t}}{\rm{. }}{\kern 8pt} \sum\nolimits_{k = 1}^K {{\rm{tr}}(\textbf{D}_m{{\textbf{{v}}_k}\textbf{{v}}_k^H}}) \le \alpha {P_m},  \forall m, \label{p10b}\\
&&  ~~~~~~~\eqref{p9b},~ \eqref{p9c},~\eqref{p9d}. \nonumber
\end{eqnarray}
\end{subequations}
{We can see that problem \eqref{p10} has semi-infinite constraints \eqref{p9b} and \eqref{p9c}, which are non-convex. To make the constraint \eqref{p9b} tractable,  the following lemma is introduced to convert \eqref{p9b} into a quadratic matrix inequality (QMI).}

\underline{\emph{\textbf{Lemma 1}}:} (Schur complement \cite{convex})  Let $ \textbf{N} $ be a complex Hermitian matrix as
\begin{eqnarray*}
\textbf{N} = \textbf{N}^{H} = \left[\begin{array}{cc}
\textbf{Y}_1 & \textbf{Y}_2 \\
\textbf{Y}_2^{H} & \textbf{Y}_3
\end{array}\right].
\end{eqnarray*}
Then, we have $ \textbf{N} \succ \textbf{0} $ if and only if $\textbf{Y}_1 - \textbf{Y}_2^H\textbf{Y}_3^{-1}\textbf{Y}_2 \succeq \textbf{0}$ with $\textbf{Y}_3 \succ \textbf{0} $, or $\textbf{Y}_3 - \textbf{Y}_2^H\textbf{Y}_1^{-1}\textbf{Y}_2 \succeq \textbf{0}$ with $\textbf{Y}_1 \succ \textbf{0}$.  ${\kern 50pt} \blacksquare$

%Then, $ \textbf{Y}_4 = \textbf{Y}_3 - \textbf{Y}_2^{H}\textbf{Y}_1^{-1}\textbf{Y}_2 $ is the Schur complement of $ \textbf{Y}_1 $ in $ \textbf{N} $, and the following statements hold:
%\begin{itemize}
%\item $ \textbf{N} \succ \textbf{0} $, if and only if $~ \textbf{Y}_1 \succ \textbf{0} $ and $ ~\textbf{Y}_4 \succ \textbf{0} $.
%\item If $~ \textbf{Y}_1 \succ \textbf{0} ~$ then $~ \textbf{N} \succ \textbf{0} ~$ if and only if $~\textbf{Y}_4 \succ \textbf{0} $.   {\kern 210pt} $\blacksquare $% %% shuanglan {\kern 20pt} $\blacksquare $
%\end{itemize}

Let us define an $MN_T\times MN_T$ square matrix ${{\textbf{V}}}_k$ as ${{{\textbf{V}}}_k}={{\textbf{v}}_{k}}{\textbf{v}}_{k}^H$. By utilizing {Lemma 1}, the constraint \eqref{p9b} can be converted into
\begin{eqnarray} \label{p11}
 \left[\begin{array}{cc}
{\zeta _k}\left( {1 - {{ \rho }_k}} \right) & \sqrt{{e}_{k}} \\
\sqrt{{e}_{k}} & \big(\hat{{\textbf{h}}}_{k} + \Delta {\textbf{h}}_{k}\big)^{H}\textbf{R} \big(\hat{{\textbf{h}}}_{k} + \Delta {\textbf{h}}_{k}\big) \!+\! \sigma_{k}^{2}
\end{array} \right] \succeq \textbf{0},
\end{eqnarray}
where $ \textbf{R} \triangleq \sum_{k=1}^{K}\textbf{V}_{k}$. Note that \eqref{p11} is still non-convex.
{In order to remove the channel uncertainty in \eqref{p11},  the following lemma is required to convert the constraint \eqref{p11} into linear matrix inequality (LMI).}

\emph{\underline{\textbf{Lemma 2}:}} \cite[Theorem 3.5]{Luo}
	Let us denote $\textbf{U}_k \in \mathbb{C}$, for $k\in[1, 6]$. If $\textbf{T}_i \!\succeq\! \textbf{0}$ for $i=1,2$, then the following QMI
	\begin{eqnarray}
	& \left[\!\!\begin{array}{cc}
	\textbf{U}_1 \!&\! \textbf{U}_2+\textbf{U}_3\textbf{X} \\
	(\textbf{U}_2\!+\!\textbf{U}_3\textbf{X})^{H} \!&\! \textbf{U}_4\!+\!\textbf{X}^{H}\textbf{U}_5\!+\!\textbf{U}_5^{H}\textbf{X}\!+\!\textbf{X}^{H}\textbf{U}_6\textbf{X}
	\end{array}
	\!\!\right] \!\succeq\! \textbf{0},\nonumber\\
	&~~~~~~  \textbf{I} \!-\! \textbf{X}^{H}\textbf{T}_i\textbf{X} \!\succeq\! \textbf{0}, ~{\textrm{for}}~\forall  \textbf{X} \nonumber
	\end{eqnarray}
 are equivalent to the LMI
	\begin{eqnarray}
    \left[\!\!\begin{array}{ccc}
	\textbf{U}_1 \!&\! \textbf{U}_2 \!&\! \textbf{U}_3 \\
	\textbf{U}_2^{H} \!&\! \textbf{U}_4 \!&\! \textbf{U}_5^{H} \\
	\textbf{U}_3^{H} \!&\! \textbf{U}_5 \!&\! \textbf{U}_6
	\end{array}
	\!\!\right] + \lambda_1\left[\!\!\begin{array}{ccc}
	\textbf{0} \!&\! \textbf{0} \!&\! \textbf{0} \\
	\textbf{0} \!&\! \textbf{I} \!&\! \textbf{0} \\
	\textbf{0} \!&\! \textbf{0} \!&\! \textbf{T}_1
	\end{array}
	\!\!\right] +  \lambda_2\left[\!\!\begin{array}{ccc}
	\textbf{0} \!&\! \textbf{0} \!&\! \textbf{0} \\
	\textbf{0} \!&\! \textbf{I} \!&\! \textbf{0} \\
	\textbf{0} \!&\! \textbf{0} \!&\! \textbf{T}_2
	\end{array}
	\!\!\right] \!\succeq\! \textbf{0}, \nonumber {\kern 95pt}
	\end{eqnarray}  %$ {\kern 80pt}  \blacksquare $ %{\kern 380pt}
where $\lambda_i \geq 0 ~(i=1,2)$. ${\kern 140pt} \blacksquare$

To proceed, we set $\textbf{X}= {\Delta {\textbf{h}}_{k} }$, $\textbf{T}_1 = 1/\varepsilon^2_{k}\textbf{I}$,  $\textbf{T}_2 = \textbf{0}$, $\textbf{U}_1 = 1-\rho_{k}$,  $\textbf{U}_2 = \sqrt{{e}_{k}}$, $\textbf{U}_3 = {\textbf{0}}_{1\times MN_T}$, $\textbf{U}_4 ={\hat{\textbf{h}}}_{k}^{H}\textbf{R}{\hat{\textbf{h}}}_{k} + \sigma_{k}^{2} - t_{k}$, $\textbf{U}_5 ={\hat{\textbf{h}}}_{k}^{H}{\textbf{R}}$, $\textbf{U}_6 =\textbf{R}$.
{Then, by exploiting Lemma 2, the constraint \eqref{p11} can be equivalently modified as the following convex LMI}
\begin{align}\label{eq:12}
\!\!\!\! \!\!\!  {{{\textbf{A}}}_k} = \left[ \begin{array}{ccc}
{\zeta _k}(1{\rm{-}}\rho_{k}) & \sqrt{{e}_{k}} & \textbf{0}_{1 \times MN_{T}} \\
\sqrt{{e}_{k}} & {\hat{\textbf{{{h}}}}}_{k}^{H}\textbf{R}{\hat{\textbf{{{h}}}}}_{k} + \sigma_{k}^{2} - t_{k} & {\hat{\textbf{{{h}}}}}_{k}^{H}{\textbf{R}} \\
\textbf{0}_{MN_{T} \times 1} & {\textbf{R}}{\hat{\textbf{{{h}}}}}_{k} & {\textbf{R}} + \frac{t_{k}}{\varepsilon_{k}^{2}}{\textbf{I}}
\end{array}
\right] \succeq \textbf{0},
\end{align}
where $t_{k} \geq 0$ is a slack variable.

{Next, we transform the constraint \eqref{p9c} to the convex one. Due to the definition of ${\rm{SIN}}{{\rm{R}}_k}$ and ${{\cal H}_k}$, the
constraint \eqref{p9c} can be recast as}
\begin{equation*} \label{p13}
\!\!\!\!\!\! {{\rho }_k}{\big| {{{\big( {{{\hat {{\textbf{h}}}}_k} \!+\! \Delta {{{\textbf{h}}_k}}} \big)}^H}{{{\textbf{v}}_k}}} \big|^2}\\
\ge {\Gamma}\big( {{{\rho }_k}\sum\limits_{j \ne k} {{{\big| {{{\big( {{{\hat {{\textbf{h}}}}_k} \!+\! \Delta {{{\textbf{h}}_k}}} \big)}^H}{{\textbf{v}}_j}} \big|}^2}}  \!+\! {{\rho }_k}\sigma _k^2 \!+\! \delta _k^2} \big),
\end{equation*}
%%% shuanglan
%\begin{equation} \label{p13}
%\begin{split}
%&{{\rho }_k}{\big| {{{\big( {{{\hat {{\textbf{h}}}}_k} + \Delta {{{\textbf{h}}_k}}} \big)}^H}{{{\textbf{v}}_k}}} \big|^2}\\
%\ge &{\Gamma}\big( {{{\rho }_k}\sum\limits_{j \ne k} {{{\big| {{{\big( {{{\hat {{\textbf{h}}}}_k} + \Delta {{{\textbf{h}}_k}}} \big)}^H}{{\textbf{v}}_j}} \big|}^2}}  + {{\rho }_k}\sigma _k^2 + \delta _k^2} \big),
%\end{split}
%\end{equation}
and thus, {it follows}
\begin{equation}\label{p14}
{\rho _k}\big({\big( {{{\hat {\textbf{h}}}_k} + \Delta {\textbf{h}}_k} \big)^H}{{{\textbf{M}}}_k} \big( {{{\hat {\textbf{h}}}_k} + \Delta {\textbf{h}}_k} \big)+\sigma _k^2 \big) \ge  {\delta _k^2},
\end{equation}
where ${{{\textbf{M}}}_k} = \frac{1}{{\Gamma}}{{{\textbf{V}}}_k}
- \sum\limits_{j \ne k} {{{{\textbf{V}}}_j}}$.

 {Also, we utilize a similar methodology for \eqref{p14} as follows. By applying Lemma 1, the constraint \eqref{p14} can be changed into}
\begin{eqnarray} \label{p15}
 \left[\begin{array}{cc}
 {{ \rho }_k}  & \delta_k \\
\delta_k  & \big(\hat{{\textbf{h}}}_{k} + \Delta {\textbf{h}}_{k}\big)^{H}{{{\textbf{M}}}_k} \big(\hat{{\textbf{h}}}_{k} + \Delta {\textbf{h}}_{k}\big) \!+\! \sigma_{k}^{2}
\end{array} \right] \succeq \textbf{0}.
\end{eqnarray}
{In order to get rid of the channel uncertainty $\Delta {\textbf{h}}_{k}$ in \eqref{p15},  Lemma 2 is adopted, and the constraint \eqref{p15} is equivalently modified as}
\begin{align}\label{p16}
\!\!\!\!\!\! {{{\textbf{B}}}_k} = \left[ \begin{array}{ccc}
\rho_{k} & \delta_k & \textbf{0}_{1 \times MN_{T}} \\
\delta_k & {\hat{\textbf{{{h}}}}}_{k}^{H}{{{\textbf{M}}}_k}{\hat{\textbf{{{h}}}}}_{k} + \sigma_{k}^{2} - r_{k} & {\hat{\textbf{{{h}}}}}_{k}^{H}{{{\textbf{M}}}_k} \\
\textbf{0}_{MN_{T} \times 1} & {{{\textbf{M}}}_k}{\hat{\textbf{{{h}}}}}_{k} & {{{\textbf{M}}}_k} + \frac{r_{k}}{\varepsilon_{k}^{2}}{\textbf{I}}
\end{array}
\right] \succeq \textbf{0},
\end{align}
where $r_{k} \geq 0$ is a slack variable.

Defining ${{\hat{\textbf{{V}}}}}_{m,k}$ as ${{\hat{\textbf{{V}}}}}_{m,k} = \textbf{D}_m{{{\textbf{V}}}}_k$, problem \eqref{p10} is thus reformulated as
\begin{equation} \label{p17}
\begin{split}
& \displaystyle \mathop {\min }\limits_{\{{{\textbf{V}}}_k\},{\kern 1pt}\rho_k,{\kern 1pt}\alpha,{\kern 1pt}{{t_k }},{\kern 1pt}{{r_k }}}  {\kern 10pt} \alpha {\kern 50pt} \\
& \displaystyle {\kern 1pt} {\rm{s}}{\rm{.t}}{\rm{. }}{\kern 6pt} \sum \nolimits_{k=1}^K {\rm{tr}}({{\hat{\textbf{{V}}}}}_{m,k})  \le \alpha {P_m}, ~\forall m,\\
& \displaystyle {\kern 25pt} {{{\textbf{A}}}_k}\succeq {\textbf{0}}, ~{{{\textbf{B}}}_k}\succeq{\textbf{0}}, ~{{{\textbf{V}}}_k}\succeq {\textbf{0}}, ~\eqref{p9d},\\
& {\kern 26pt} {t _k} \ge 0, ~{ r_k} \ge 0,  ~{\rm{rank}}({{{\textbf{V}}}_k}) = 1,\forall k.
\end{split}
\end{equation}
%% shuanglan
%\begin{equation} \label{p17}
%\begin{split}
%& \displaystyle \mathop {\min }\limits_{\{{{\textbf{V}}}_k\},{\kern 1pt}\rho_k,{\kern 1pt}\alpha,{\kern 1pt}{{t_k }},{\kern 1pt}{{r_k }}}  {\kern 10pt} \alpha {\kern 50pt} \\
%& \displaystyle {\kern 1pt} {\rm{s}}{\rm{.t}}{\rm{. }}{\kern 1pt}{\kern 1pt} \sum \nolimits_{k=1}^K {\rm{tr}}({{\hat{\textbf{{V}}}}}_{m,k})  \le \alpha {P_m}, \forall m,\\
%& \displaystyle {\kern 18pt} {{{\textbf{A}}}_k}\succeq {\textbf{0}}, ~{{{\textbf{B}}}_k}\succeq{\textbf{0}}, ~{{{\textbf{V}}}_k}\succeq {\textbf{0}},~0 < {\rho _k} \leq 1,\\
%& {\kern 21pt} {t _k} \ge 0,{ r_k} \ge 0, \eqref{p9d},{\rm{rank}}({{{\textbf{V}}}_k}) = 1,\forall k.
%\end{split}
%\end{equation}

The above optimization problem is difficult to solve in general due to the rank-one constraint. Therefore, we employ the semi-definite relaxation (SDR) technique \cite{SDR} which simply drops the constraints ${\rm{rank}}({{{\textbf{V}}}_k}) = 1$ for all ${{{\textbf{V}}}_k}$'s. Then, problem \eqref{p17} becomes a convex problem which can be solved efficiently by a convex programming solver such as CVX \cite{CVX}.
%As mentioned before, problem (14) may not give the optimal solution to the original problem (10) due to the rank relaxation. Let $\left\{ {{{\textbf{X}}}_k^*} \right\}$ denote the optimal solution of problem (14) with a given ${\rho _k}$. If $\left\{ {{{\textbf{X}}}_k^*} \right\}$ satisfies ${\rm{rank}}({{\textbf{X}}}_k^*) = 1$, then the optimal robust transmit beamforming solution ${\textbf{v}}_k^*$ for problem (9) can
%be obtained by eigenvalue decomposition (EVD) on ${{\textbf{X}}}_k^*$.
In the following theorem, we show that a solution ${{{\textbf{V}}}_k^*}$ to problem \eqref{p17} satisfies ${\rm{rank}}({{\textbf{V}}}_k^*) \leq 2$.

{{\underline{\emph{\textbf{Theorem 1}}}:}} If problem \eqref{p17} is feasible, the rank of a solution ${{{\textbf{V}}}_k^*}$ to problem \eqref{p17} via rank relaxation  is less than or equal to 2.

{{\emph{\underline{Proof}:}}} See Appendix A.  $ {\kern 131pt} \blacksquare$ %% %shaunglan  $ {\kern 130pt} \blacksquare$

%The optimal  transmit beamforming ${\textbf{v}}_k^*$ for problem \eqref{p9} can be obtained by applying eigenvalue decomposition (EVD) on ${{\textbf{X}}}_k^*$.

After ${{\textbf{V}}}_k^*$ is obtained, if rank$({{\textbf{V}}}_k^*) = 1$, we can compute an optimal transmit beamforming solution ${\textbf{v}}_{k}$ by eigenvalue decomposition (EVD) of ${{\textbf{V}}}_k^*$. If rank$({{\textbf{V}}}_k^*) = 2$, we use the conventional Gaussian randomization (GR) technique \cite{SDR} to find  ${\textbf{v}}_{k}$ for $k = 1,...,K$. In particular, the GR technique generates a suboptimal solution. Hence, when rank$({{\textbf{V}}}_k^*) = 2$, we will propose an iterative algorithm to recover the optimal rank-one solution by following the approach in \cite{penalty_function}.

First, since ${{\hat{\textbf{{V}}}}}_{m,k}$ is always semi-positive definite, we have ${\rm{tr}}({{\hat{\textbf{{V}}}}}_{m,k}) \geq \lambda_{{\rm{max}}}({{\hat{\textbf{{V}}}}}_{m,k})$. Thus,  we can prove that rank$({{\hat{\textbf{{V}}}}}_{m,k}) = 1$ if ${\rm{tr}}({{\hat{\textbf{{V}}}}}_{m,k}) \leq \lambda_{{\rm{max}}}({{\hat{\textbf{{V}}}}}_{m,k})$. Then, we can transform {the constraint rank$({{\hat{\textbf{{V}}}}}_{m,k}) = 1$}
into the single reverse convex constraint as
\begin{equation*}\label{p17_1}
\sum\limits_{k = 1}^K ({{\rm{tr}}({{\hat{\textbf{{V}}}}}_{m,k}}) - \lambda_{{\rm{max}}}({{\hat{\textbf{{V}}}}}_{m,k})) \le 0.
\end{equation*}
Note that the function $\lambda_{{\rm{max}}}({{\hat{\textbf{{V}}}}}_{m,k})$ on the set of Hermitian
matrices is convex. When $\sum\limits_{k = 1}^K ({{\rm{tr}}({{\hat{\textbf{{V}}}}}_{m,k}}) - \lambda_{{\rm{max}}}({{\hat{\textbf{{V}}}}}_{m,k}))$ is small enough,
${{\hat{\textbf{{V}}}}}_{m,k}$ will approach $\lambda_{{\rm{max}}}({{\hat{\textbf{{V}}}}}_{m,k})\hat{\textbf{{v}}}^{{\rm{max}}}_{m,k}(\hat{\textbf{{v}}}^{{\rm{max}}}_{m,k})^H$, where $\hat{\textbf{{v}}}^{{\rm{max}}}_{m,k}$ represents the eigenvector corresponding to the maximum eigenvalue $\lambda_{{\rm{max}}}({{\hat{\textbf{{V}}}}}_{m,k})$ with $\|\hat{\textbf{{v}}}^{{\rm{max}}}_{m,k}\|=1$. Then the optimal transmit beamformer vector can be expressed by
\begin{equation}\label{p17_2}
\textbf{{v}}_{m,k} = \sqrt{\lambda_{{\rm{max}}}({{\hat{\textbf{{V}}}}}_{m,k})}\hat{\textbf{{v}}}^{{\rm{max}}}_{m,k},
\end{equation}
which satisfies the rank-one constraint.

{Thus, in order to make $\sum\limits_{k = 1}^K ({{\rm{tr}}({{\hat{\textbf{{V}}}}}_{m,k}}) - \lambda_{{\rm{max}}}({{\hat{\textbf{{V}}}}}_{m,k})) $ as small as possible, we adopt the exact penalty method \cite{convex}.
First,} introducing a sufficiently large penalty ratio $\theta > 0$, the alternative formulation is considered as
\begin{subequations}\label{p17_3}
\begin{eqnarray}
&&\!\!\!\!\!\!\!\!\!\!\!\!\!\! \mathop {\min }\limits_{\{{{\textbf{V}}}_k\},{\kern 1pt}\rho_k,{\kern 1pt}\alpha,{\kern 1pt}{{t_k }},{\kern 1pt}{{r_k }}}  {\kern 10pt} \alpha {\kern 50pt} \label{p17_3a}\\
&&\!\!\!\!\!\!\!\!\!\!\!\!\!\!  {\rm{s}}{\rm{.t}}{\rm{. }} {\kern 6pt} {{{\textbf{A}}}_k}\succeq {\textbf{0}}, ~{{{\textbf{B}}}_k}\succeq{\textbf{0}}, ~{{{\textbf{V}}}_k}\succeq {\textbf{0}},~\eqref{p9d}, \label{p17_3b}\\
&&\!\!\!\!\!\!\!\!\!\!\!\!\!\!\!\!  \sum\limits_{k = 1}^K ({{\rm{tr}}({{\hat{\textbf{{V}}}}}_{m,k}})  + \theta({{\rm{tr}}({{\hat{\textbf{{V}}}}}_{m,k}}) - \lambda_{{\rm{max}}}({{\hat{\textbf{{V}}}}}_{m,k}))) \le \alpha {P_m},\label{p17_3c}\\
&&  {t _k} \ge 0,{ r_k} \ge 0,  ~\forall k. \label{p17_3d}
\end{eqnarray}
\end{subequations}
%%% shuanglan
%\begin{subequations}\label{p17_3}
%\begin{eqnarray}
%&&\!\!\!\!\!\!\!\!\!\!\!\!\!\!\!\! \mathop {\min }\limits_{\{{{\textbf{V}}}_k\},{\kern 1pt}\rho_k,{\kern 1pt}\alpha,{\kern 1pt}{{t_k }},{\kern 1pt}{{r_k }}}  {\kern 10pt} \alpha {\kern 50pt} \label{p17_3a}\\
%&&\!\!\!\!\!\!\!\!\!\!\!\!\!\!\!\!  {\rm{s}}{\rm{.t}}{\rm{. }} {\kern 3pt} {{{\textbf{A}}}_k}\succeq {\textbf{0}}, ~{{{\textbf{B}}}_k}\succeq{\textbf{0}}, ~{{{\textbf{V}}}_k}\succeq {\textbf{0}},~0 < {\rho _k} \leq 1, \label{p17_3b}\\
%&&\!\!\!\!\!\!\!\!\!\!\!\!\!\!\!\!\!\!\!\! \sum\limits_{k = 1}^K ({{\rm{tr}}({{\hat{\textbf{{V}}}}}_{m,k}})  + \theta({{\rm{tr}}({{\hat{\textbf{{V}}}}}_{m,k}}) - \lambda_{{\rm{max}}}({{\hat{\textbf{{V}}}}}_{m,k}))) \le \alpha {P_m},\label{p17_3c}\\
%&&\!\!\!\!\!\!\!\! {\kern 3pt} {t _k} \ge 0,{ r_k} \ge 0, \eqref{p9d},\forall k. \label{p17_3d}
%\end{eqnarray}
%\end{subequations}
We can find from \eqref{p17_3c} that the difference ${{\rm{tr}}({{\hat{\textbf{{V}}}}}_{m,k}}) - \lambda_{{\rm{max}}}({{\hat{\textbf{{V}}}}}_{m,k})$ will be
minimized when $\theta$ is large
enough. Clearly, \eqref{p17_3c} is set to minimize ${{\rm{tr}}({{\hat{\textbf{{V}}}}}_{m,k}}) - \lambda_{{\rm{max}}}({{\hat{\textbf{{V}}}}}_{m,k})$.
{Note that \eqref{p17_3c} is non-convex due to the
coupled  $\theta $ and ${{\hat{\textbf{{V}}}}}_{m,k}$.
 To eliminate the coupling between $\theta $ and ${{\hat{\textbf{{V}}}}}_{m,k}$, we apply the  following lemma
to provide an effective approximation of \eqref{p17_3c}.}

%%% old %%%%%
%{{\underline{\emph{\textbf{Lemma 3}}}:}} Let us define $\textbf{C} \in \mathbb{H}_{+}$ and $\textbf{E}\in \mathbb{H}_{+}$.
%Applying the fact that a sub-gradient of $\lambda_{{\rm{max}}}(\textbf{E})$ is $\textbf{\emph{e}}_{{\rm{max}}}\textbf{\emph{e}}_{{\rm{max}}}^H$, where  $\textbf{\emph{e}}_{{\rm{max}}}$ denote the maximum eigenvalue corresponding eigenvector of $\textbf{E}$.
% It always follows $\lambda_{{\rm{max}}}(\textbf{C}) - \lambda_{{\rm{max}}}(\textbf{E})\geq \textbf{\emph{e}}_{{\rm{max}}}^H(\textbf{C} - \textbf{E})\textbf{\emph{e}}_{{\rm{max}}}$.  {\kern 215pt} $\blacksquare $
%
{{\underline{\emph{\textbf{Lemma 3}}}:}} Let us define $\textbf{C} \in \mathbb{H}_{+}$ and $\textbf{E}\in \mathbb{H}_{+}$.
Then, it always follows $\lambda_{{\rm{max}}}(\textbf{C}) - \lambda_{{\rm{max}}}(\textbf{E})\geq \textbf{\emph{e}}_{{\rm{max}}}^H(\textbf{C} - \textbf{E})\textbf{\emph{e}}_{{\rm{max}}}$, where  $\textbf{\emph{e}}_{{\rm{max}}}$ denotes the  eigenvector  corresponding to the maximum eigenvalue of $\textbf{E}$.  {\kern 131pt} $\blacksquare $

According to Lemma 3, we propose an iterative algorithm to recover a local optimal solution. For given some feasible $\{\hat{\textbf{V}}_{m,k}^{(n)}\}$ to problem \eqref{p17_3}, we get
\begin{equation}\label{p17_4}
\begin{split}
& {{\rm{tr}}({{\hat{\textbf{{V}}}}}_{m,k}^{(n+1)}})+ \theta \bigg[ {{\rm{tr}}({{\hat{\textbf{{V}}}}}_{m,k}^{(n+1)}}) - \lambda_{{\rm{max}}}({{\hat{\textbf{{V}}}}}_{m,k}^{(n)})   \\
 &-(\hat{\textbf{{v}}}_{m,k}^{{{\rm{max}}},(n)})^H\big({{\hat{\textbf{{V}}}}}_{m,k}^{(n+1)}-{{\hat{\textbf{{V}}}}}_{m,k}^{(n)}\big)
\hat{\textbf{{v}}}_{m,k}^{{{\rm{max}}},(n)}\bigg]  \\
\leq &{\kern 1pt} {{\rm{tr}}({{\hat{\textbf{{V}}}}}_{m,k}^{(n)}})  + \theta \big( {{\rm{tr}}({{\hat{\textbf{{V}}}}}_{m,k}^{(n)}}) - \lambda_{{\rm{max}}}({{\hat{\textbf{{V}}}}}_{m,k}^{(n)}) \big),
\end{split}
\end{equation}
%%% shuanglan
%\begin{equation}\label{p17_4}
%\begin{split}
%&{{\rm{tr}}({{\hat{\textbf{{V}}}}}_{m,k}^{(n+1)}})  + \theta \bigg[ {{\rm{tr}}({{\hat{\textbf{{V}}}}}_{m,k}^{(n+1)}}) - \lambda_{{\rm{max}}}({{\hat{\textbf{{V}}}}}_{m,k}^{(n)}) \\
%& -(\hat{\textbf{{v}}}_{m,k}^{{{\rm{max}}},(n)})^H\big({{\hat{\textbf{{V}}}}}_{m,k}^{(n+1)}-{{\hat{\textbf{{V}}}}}_{m,k}^{(n)}\big)
%\hat{\textbf{{v}}}_{m,k}^{{{\rm{max}}},(n)}\bigg]  \\
%&\leq {{\rm{tr}}({{\hat{\textbf{{V}}}}}_{m,k}^{(n)}})  + \theta \big( {{\rm{tr}}({{\hat{\textbf{{V}}}}}_{m,k}^{(n)}}) - \lambda_{{\rm{max}}}({{\hat{\textbf{{V}}}}}_{m,k}^{(n)}) \big),
%\end{split}
%\end{equation}
where the superscript $n$ represents the $n$-th iteration.

Hence, the following SDP problem generates an optimal solution
${{{\textbf{V}}}_{m,k}^{(n+1)}}$ that is better than ${{{\textbf{V}}}_{m,k}^{(n)}}$ to problem \eqref{p17_3} as
\begin{subequations}\label{p17_51}
\begin{eqnarray}
&& \!\!\!\!\!\!\!\mathop {\min }\limits_{\{{{\textbf{V}}}_k\},{\kern 1pt}\rho_k,{\kern 1pt}\alpha,{\kern 1pt}{{t_k }},{\kern 1pt}{{r_k }}}  {\kern 30pt} \alpha {\kern 30pt} \label{p17_5a}\\
&& \!\!\!\!\!\!\! {\rm{s}}{\rm{.t}}{\rm{. }} ~ \eqref{p17_3b}, ~\eqref{p17_3d},\\
&& \!\!\!\!\!\!\!\!\!\!\!\!\! \sum_{k=1}^{K}\bigg\{ {{\rm{tr}}({{\hat{\textbf{{V}}}}}_{m,k}})  + \theta \bigg[ {{\rm{tr}}({{\hat{\textbf{{V}}}}}_{m,k}}) - \lambda_{{\rm{max}}}({{\hat{\textbf{{V}}}}}_{m,k}^{(n)}) \nonumber  \\
&&\!\!\!\!\!\!\!\!\!\!\!\! - (\hat{\textbf{{v}}}_{m,k}^{{{\rm{max}}},(n)})^H\big({{\hat{\textbf{{V}}}}}_{m,k}-{{\hat{\textbf{{V}}}}}_{m,k}^{(n)}\big)
\hat{\textbf{{v}}}_{m,k}^{{{\rm{max}}},(n)}\bigg] \bigg\} \leq \alpha {P_m}. \label{p17_5c}
%&& \!\!\!\!\!\!\!\!\!\!\!\!\!\!\!\!\!\!\!\!\!\!\!\!\!\!\!\!\!\!\!\!\!\!\! \sum_{k=1}^{K}\bigg\{ {{\rm{tr}}({{\hat{\textbf{{V}}}}}_{m,k}^{(n+1)}})  + \theta \bigg[ {{\rm{tr}}({{\hat{\textbf{{V}}}}}_{m,k}^{(n+1)}}) - \lambda_{{\rm{max}}}({{\hat{\textbf{{V}}}}}_{m,k}^{(n)}) -   (\hat{\textbf{{v}}}_{m,k}^{{{\rm{max}}},(n)})^H\big({{\hat{\textbf{{V}}}}}_{m,k}^{(n+1)}-{{\hat{\textbf{{V}}}}}_{m,k}^{(n)}\big)
%\hat{\textbf{{v}}}_{m,k}^{{{\rm{max}}},(n)}\bigg] \bigg\} \leq \alpha {P_m}. \label{p17_5c}
\end{eqnarray}
\end{subequations}
%% shuanglan
%\begin{subequations}\label{p17_5}
%\begin{eqnarray}
%&&\!\!\!\!\!\!\!\!\!\!\!\!\!\!\!\! \mathop {\min }\limits_{\{{{\textbf{V}}}_k\},{\kern 1pt}\rho_k,{\kern 1pt}\alpha,{\kern 1pt}{{t_k }},{\kern 1pt}{{r_k }}}  {\kern 30pt} \alpha {\kern 30pt} \label{p17_5a}\\
%&&\!\!\!\!\!\!\!\!\!\!\!\!\!\!\!\!  {\rm{s}}{\rm{.t}}{\rm{. }} ~~ \eqref{p17_3b}, ~\eqref{p17_3d},\\
%&&\!\!\!\!\!\!\!\!\!\!\!\!\!\!\!\! \sum_{k=1}^{K}\bigg\{ {{\rm{tr}}({{\hat{\textbf{{V}}}}}_{m,k}^{(n+1)}})  + \theta \bigg[ {{\rm{tr}}({{\hat{\textbf{{V}}}}}_{m,k}^{(n+1)}}) - \lambda_{{\rm{max}}}({{\hat{\textbf{{V}}}}}_{m,k}^{(n)}) - \nonumber \\
%&&\!\!\!\!\!\!\!\!\!\!\!\!\!\!\!\!  (\hat{\textbf{{v}}}_{m,k}^{{{\rm{max}}},(n)})^H\big({{\hat{\textbf{{V}}}}}_{m,k}^{(n+1)}-{{\hat{\textbf{{V}}}}}_{m,k}^{(n)}\big)
%\hat{\textbf{{v}}}_{m,k}^{{{\rm{max}}},(n)}\bigg] \bigg\} \leq \alpha {P_m}, \forall m. \label{p17_5c}
%\end{eqnarray}
%\end{subequations}
Now, problem \eqref{p17_51} can be further simplified to
\begin{subequations}\label{p17_5}
\begin{eqnarray}
&&\!\!\!\!\!\!\!\!\!\!\!\!\!\!\!\! \mathop {\min }\limits_{\{{{\textbf{V}}}_k\},{\kern 1pt}\rho_k,{\kern 1pt}\alpha,{\kern 1pt}{{t_k }},{\kern 1pt}{{r_k }}}  {\kern 30pt} \alpha {\kern 30pt} \label{p17_5a}\\
&&\!\!\!\!\!\!\!\!\!\!\!\!\!\!\!\!  {\rm{s}}{\rm{.t}}{\rm{. }} ~~ \eqref{p17_3b}, ~\eqref{p17_3d},\\
&&\!\!\!\! \sum_{k=1}^{K}\bigg\{ {{\rm{tr}}({{\hat{\textbf{{V}}}}}_{m,k}})  + \theta \big[{{\rm{tr}}({{\hat{\textbf{{V}}}}}_{m,k}}) \nonumber \\
&&\!\!\!\! -(\hat{\textbf{{v}}}_{m,k}^{{{\rm{max}}},(n)})^H{{\hat{\textbf{{V}}}}}_{m,k}\hat{\textbf{{v}}}_{m,k}^{{{\rm{max}}},(n)}\big]\bigg\} \leq \alpha {P_m}, \forall m.\label{p17_5c}
\end{eqnarray}
\end{subequations}

To summarize, we can solve problem \eqref{p8} with a given $\Gamma$, and a bisection search algorithm is applied to update $\Gamma$ for the objective value ${\alpha }^*=1$. Then, this process is repeated until convergence. For the bisection method, we need to determine an upper bound ${\Gamma_{{\mathop{\rm m}\nolimits} {\rm{ax}}}}$ as $0 < {\Gamma}<{\Gamma_{{\mathop{\rm m}\nolimits} {\rm{ax}}}}$. Then,  we can see that
%\begin{equation*}
%\begin{split}
% {{\rm{SIN}}{{\rm{R}}_k}\left( {{\{\textbf{{v}}_{k}\}},{{ \rho }_k}} \right)} &=\frac{{{{ \rho }_k}{{| {{\textbf{h}}_{k}^H{{\textbf{v}}_{k}}} |}^2}}}{{ {{{ \rho }_k} {{ \sum\limits_{j \ne k}{| {{\textbf{h}}_{k}^H{{\textbf{v}}_{j}}} |}^2}}  + {{ \rho }_k}\sigma _k^2 + \delta _k^2} }}  \le \frac{{{{ \rho }_k}{{| {{\textbf{h}}_{k}^H{{\textbf{v}}_{k}}} |}^2}}}{{ {{{ \rho }_k}\sigma _k^2 + \delta _k^2} }} \\
%& \le \frac{{{{ \rho }_k}{{ \left\| {{\textbf{h}}_{k}} \right\|}^2} \sum_{j=1}^M P_m }}{{ {{{ \rho }_k}\sigma _k^2 + \delta _k^2} }} \le \frac{{{{ \left\| {{\textbf{h}}_{k}} \right\|}^2} \sum_{j=1}^M P_m  }}{{ {\sigma _k^2 + \delta _k^2} }}.
%\end{split}
%\end{equation*}
%% shuanglan
\begin{equation*}
\begin{split}
& {{\rm{SIN}}{{\rm{R}}_k}\left( {{\{\textbf{{v}}_{k}\}},{{ \rho }_k}} \right)} =\frac{{{{ \rho }_k}{{| {{\textbf{h}}_{k}^H{{\textbf{v}}_{k}}} |}^2}}}{{ {{{ \rho }_k} {{ \sum\limits_{j \ne k}{| {{\textbf{h}}_{k}^H{{\textbf{v}}_{j}}} |}^2}}  + {{ \rho }_k}\sigma _k^2 + \delta _k^2} }}\\
& ~~~ \le \frac{{{{ \rho }_k}{{| {{\textbf{h}}_{k}^H{{\textbf{v}}_{k}}} |}^2}}}{{ {{{ \rho }_k}\sigma _k^2 + \delta _k^2} }} \le \frac{{{{ \rho }_k}{{ \left\| {{\textbf{h}}_{k}} \right\|}^2} \sum_{j=1}^M P_m }}{{ {{{ \rho }_k}\sigma _k^2 + \delta _k^2} }}  \le \frac{{{{ \left\| {{\textbf{h}}_{k}} \right\|}^2} \sum_{j=1}^M P_m  }}{{ {\sigma _k^2 + \delta _k^2} }}.
\end{split}
\end{equation*}
From this, we can set
${\Gamma_{\max }}$  as $\mathop {\max }\limits_k \left\{ \frac{{{{ \left\| {{\textbf{h}}_{k}} \right\|}^2}  \sum_{j=1}^M P_m  }}{{ {\sigma _k^2 + \delta _k^2} }} \right\}$. Due to monotonicity of ${\alpha }$, the bisection search algorithm needs ${\cal O}\left(
{{{\log }_2} {\frac{{{\Gamma_{\max }} }}{\eta }} } \right)$ iterations, where $\eta $ is a small positive constant which controls the accuracy of the bisection search algorithm. It is noted that this bisection search algorithm converges to the optimal solution ${\textbf{v}}_{k}^*$ for problem \eqref{p8}. The proposed algorithm based on bisection search is summarized in Algorithm~1.\footnote{{The proposed optimization algorithm is performed by MPU. Then, the MPU can send the beamforming solutions to individual transmitters through fiber optics or an exclusive radio frequency (RF) link. Also, it can transmit the PS factor solution to individual receivers through the estimated instantaneous channel.}}

\begin{algorithm}
\caption{ Proposed algorithm based on bisection search}
\label{alg: }
\begin{algorithmic}
\STATE   {Set $\displaystyle  {\Gamma_{\min }} = 0$, ${\Gamma_{\max }} = {\max \limits_k}\left\{ {\frac{{{{ \left\| {{\textbf{h}}_{k}} \right\|}^2}\sum_{j=1}^M P_m }}{{{\sigma_k^2 +
\delta _k^2} }}} \right\}$, $n = 0$, $\theta > 0$, a prescribed accuracy tolerance $\epsilon>0$ and $\eta >0$.  Randomly generate an initial value $\left\{ \textbf{V}_k^{(0)}, \rho_k^{(0)}  \right\}, \forall k$ in \eqref{p17_5}.}

\textbf{Repeat}

${\kern 10pt} $  Set $\Gamma_{\rm{mid}} = (\Gamma_{\min} +\Gamma_{\max})/2$.

\STATE  ${\kern 10pt} $ \textbf{Repeat}

 \STATE ${\kern 25pt}$  Solve problem \eqref{p17_5} with $\Gamma_{\rm{mid}}$ to obtain  a solution ${{\textbf{V}}}_k^{(n+1)}$ and  $\rho_k^{(n+1)}$.
  %% shuanglan $~~~~~~~~~~~~{{\textbf{X}}}_k^{(n+1)}$ and $\rho_k^{(n+1)}$;

 \STATE ${\kern 25pt}$  If ${{\hat{\textbf{{V}}}}}_{m,k}^{(n+1)} = {{\hat{\textbf{{V}}}}}_{m,k}^{(n)}$, set $\theta \leftarrow 2\theta$.

 \STATE ${\kern 25pt}$  Update n $\leftarrow$ n+1.

 \STATE  ${\kern 10pt} $ \textbf{Until} $|{\rm{tr}}({{\hat{\textbf{{V}}}}}_{m,k}^{(n)}) - \lambda _{\max}({{\hat{\textbf{{V}}}}}_{m,k}^{(n)})| < \epsilon$

 \STATE ${\kern 10pt} $  Set ${{\textbf{V}}}_k^{(0)} = {{\textbf{V}}}_k^{(n)}$, $\rho_k^{(0)}=\rho_k^{(n)}$, and $n =0$.

 \STATE ${\kern 10pt} $ \textbf{Repeat}

 \STATE ${\kern 25pt}$  Solve problem \eqref{p17_5} with $\Gamma_{\rm{mid}}$ to obtain  a solution ${{\textbf{V}}}_k^{(n+1)}$, $\rho_k^{(n+1)}$, and $\alpha^{(n+1)}$.

  %% shaunglan  $~~~~~~~~~~~~~{{\textbf{X}}}_k^{(n+1)}$, $\rho_k^{(n+1)}$ and $\alpha^{(n+1)}$. Denote $\alpha^{(n+1)} = \alpha^*$;

 \STATE ${\kern 25pt}$  Update n $\leftarrow$ n+1.

  \STATE  ${\kern 10pt} $ \textbf{Until} $|{\rm{tr}}({{\hat{\textbf{{V}}}}}_{m,k}^{(n)}) - \lambda _{\max}({{\hat{\textbf{{V}}}}}_{m,k}^{(n)})|<\epsilon$

 \STATE ${\kern 10pt} $ If $\alpha^{(n+1)} < 1$,

${\kern 25pt}$ set $\Gamma_{\min} = \Gamma_{\rm{mid}}$.

\STATE ${\kern 10pt}$  else

${\kern 25pt}$  set $\Gamma_{\max} =$ $\Gamma_{\rm{mid}}$.

 \STATE   \textbf{Until}  $|{\Gamma_{\max} -  \Gamma_{\min}} |<\eta$

Calculate $\textbf{{v}}_k$ according to \eqref{p17_2}.

\end{algorithmic}
\end{algorithm}

\subsection{Robust Iterative Algorithm Based on Successive Convex Approximation}
To reduce the computational complexity of Algorithm 1, we consider another formulation for the minimum SINR maximization problem. Based on the SCA method, the optimization can also be reformulated into a convex form with linear constraints. Thus, the robust SINR maximization problem can be rewritten as
\begin{subequations}\label{p18}
\begin{eqnarray}
&& \!\!\!\!\!\!\!\!\!\!\!\!\!\!\!\!\!\!\!\!\!\!\! \mathop {\min }\limits_{{\{{{\textbf{v}}}_{k}\}, { {\kern 1pt} {{\rho _k}} }}} {\kern 5pt} \mathop {\max }\limits_{{\textbf{h}_k} \in {{\cal H}_k}}   {\kern 7pt} \frac{{{{| {{\textbf{h}}_k^H{{\textbf{v}}_k}}|}^2}}}{{\sum\limits_{j \ne k} {{{| {{\textbf{h}}_k^H{{\textbf{v}}_j}} |}^2}}  + \sigma _k^2 + \frac{\delta _k^2}{{\rho _k}}}} \label{p18a} \\
&& \!\!\!\!\!\!\!\!\!\!\!\!\!\!\!\!\!\!\!\!\!\!\!  {\rm{s.t.}} ~ \mathop {\min }\limits_{{\textbf{h}_k} \in {{\cal H}_k}} {\zeta _k}\left( {1 - {\rho _k}} \right)( { \sum\nolimits_{j = 1}^K {{{| {{{\textbf{h}}}_{k}^H{{{\textbf{v}}}_{j}}} |}^2} + \sigma _k^2} } ) \ge {e_k}, ~\forall k, \label{p18b}\\
&& \!\!\!\!\!\!\!\!\!\!\!\!\!\!\!\!\!\!\!\!\!\!\! ~~~~~ \eqref{p8c},~\eqref{p8d}.\nonumber
\end{eqnarray}
\end{subequations}
%% shuanglan
%\begin{subequations}\label{p18}
%\begin{eqnarray}
%&&\!\!\!\!\!\!\!\!\!\!\!\!\!\!\!\!  \mathop {\max }\limits_{{\{{{\textbf{v}}}_{k}\}, { {\kern 1pt} {{\rho _k}} }}} {\kern 5pt} \mathop {\min }\limits_{{\textbf{h}_k} \in {{\cal H}_k}}   {\kern 7pt} \frac{{{{| {{\textbf{h}}_k^H{{\textbf{v}}_k}}|}^2}}}{{\sum\limits_{j \ne k} {{{| {{\textbf{h}}_k^H{{\textbf{v}}_j}} |}^2}}  + \sigma _k^2 + \frac{\delta _k^2}{{\rho _k}}}} \label{p18a} \\
%&&\!\!\!\!\!\!\!\!\!\!\!\!\!\!\!\!  {\rm{s.t.}} \mathop {\min }\limits_{{\textbf{h}_k} \in {{\cal H}_k}} {\zeta _k}\left( {1 - {\rho _k}} \right)( { \sum\nolimits_{j = 1}^K {{{| {{{\textbf{h}}}_{k}^H{{{\textbf{v}}}_{j}}} |}^2} + \sigma _k^2} } ) \ge {e_k}, \label{p18b}\\
%&&\!\!\!\!\!  \eqref{p8c},~\eqref{p8d}.\nonumber
%\end{eqnarray}
%\end{subequations}

In this problem, we minimize the numerator of SINR while maximizing the denominator of SINR \cite{zhu_16JCN_SWIPT_PS}. {Based on a tight approximation, the minimum and the maximum for each term can be determined by employing the Lagrangian multiplier method. In addition, to equivalently convert the objective function \eqref{p18a}, we introduce the exponential variables $ e^{x_k} $ and $e^{y_k}$ as }
\begin{subequations}\label{p19}
\begin{eqnarray}
&& \displaystyle e^{x_k} \leq \mathop {\min }\limits_{{\textbf{h}_k} \in {{\cal H}_k}} {{| {{\textbf{h}}_k^H{{\textbf{v}}_k}}|}^2}, \label{p19a} \\
&& \displaystyle e^{y_{{k}}} \geq \mathop {\max }\limits_{{\textbf{h}_k} \in {{\cal H}_k}}  \sum\limits_{j \ne k} {{{| {{\textbf{h}}_k^H{{\textbf{v}}_j}} |}^2}}  + \sigma _k^2 + \frac{\delta _k^2}{{\rho _k}}. \label{p19b}
\end{eqnarray}
\end{subequations}
Thus, {in order to circumvent the non-convex objective function \eqref{p18a},} problem \eqref{p18} is expressed by introducing a slack variable $\tau$ as
\begin{subequations} \label{p20}
\begin{eqnarray}
&& \displaystyle
\mathop {\min }\limits_{{\{{{\textbf{v}}}_{k}\}, { {\kern 1pt} {{\rho _k}} }, {\kern 1pt} \tau, {\kern 1pt}x_k, {\kern 1pt}{y_{{k}}}}}    {\kern 10pt} \tau \nonumber \\
&& \displaystyle {\rm{s.t.} } ~ e^{x_k - {y_{{k}}}} \leq \tau, \label{p20a}\\
&& ~~~~ {\kern 1pt}\eqref{p8c}, {\kern 1pt}\eqref{p8d},{\kern 1pt}\eqref{p18b},  {\kern 1pt}\eqref{p19a},{\kern 1pt}\eqref{p19b}.
\end{eqnarray}
\end{subequations}

Note that \eqref{p19b} is in concave form. Defining  $y_k^{(n)}$ as the variables $y_k$ at the $n$-th iteration for an SCA
iterative algorithm, a Taylor series expansion  ${e^{{z_k^{(n)}}}}({z_k} - z_k^{(n)} + 1) \le {e^{{z_k}}}$ is adopted to linearize \eqref{p19b}  as
\begin{eqnarray}
e^{{y}_{{k}}^{(n)}}({{y}_{{k}}} {\rm{-}} {{y}_{{k}}^{(n)}}{\rm{+}}1) \geq \mathop {\max }\limits_{{\textbf{h}_k} \in {{\cal H}_k}}  \sum\limits_{j \ne k} {{{| {{\textbf{h}}_k^H{{\textbf{v}}_j}} |}^2}}  + \sigma _k^2 + \frac{\delta _k^2}{{\rho _k}}.\label{p19_1}
\end{eqnarray}
%where ${{\hat{y}}_{{k}}}$ is an approximated value such that ${{{y}}_{{k}}} = {{\hat{y}}_{{k}}}$
%when the approximation is tight.
{When computing the EH constraint in \eqref{p18b} and the SINR constraint in \eqref{p19_1}, we need to calculate ${| {{{{\textbf{h}}}}_k^H{{{{\textbf{v}}}}_j}} |^2}$.
%Defining ${{{\textbf{{V}}}}_j} \triangleq {{{{\textbf{v}}}}_j}{{{\textbf{v}}}}_j^H$ and
Using ${{\textbf{x}}^H}{\rm{\textbf{A}}}{\textbf{x}}{\rm{ = tr}}\left( {{\rm{\textbf{A}}}{\textbf{x}}{{\textbf{x}}^H}} \right)$, we can write this as}
{\begin{equation*} \label{p21}
\begin{split}
\displaystyle {| {{{{\textbf{h}}}}_k^H{{{{\textbf{v}}}}_j}} |^2} & ={| {{{( {{{{{\hat {{\textbf{h}}}}}}_k} + \Delta {{\emph{{\textbf{h}}}}_k}} )}^H}{{{{\textbf{v}}}}_j}} |^2}\\
&={{{\textbf{v}}}}_j^H( {{{{{\hat {{\textbf{h}}}}}}_k} + \Delta {{\emph{{\textbf{h}}}}_k}} ){( {{{{{\hat {{\textbf{h}}}}}}_k} + \Delta {{\emph{{\textbf{h}}}}_k}} )^H}{{{{\textbf{v}}}}_j}\\
& = {\rm{tr}}\big( {( {{{{{\hat {{\textbf{h}}}}}}_k} + \Delta {{\emph{{\textbf{h}}}}_k}} ){{( {{{{{\hat {{\textbf{h}}}}}}_k} + \Delta {{\emph{{\textbf{h}}}}_k}} )}^H}{{{{\textbf{v}}}}_j}{{{\textbf{v}}}}_j^H} \big)\\
& = {\rm{tr}}\big( {( {{{{{\hat {{\textbf{H}}}}}}_k} + {\Delta _k}} ){{{\textbf{{V}}}}_j}} \big)
\end{split}
\end{equation*}
where ${\hat {\textbf{H}}_k}$ is defined as ${\hat {\textbf{H}}_k}\triangleq{\hat {{\textbf{h}}}_k}\hat {{\textbf{h}}}_k^H$, and ${\Delta _k}\triangleq {\hat {{\textbf{h}}}_k}\Delta {{\textbf{h}}}_k^H + \Delta {{\textbf{h}}_k}\hat {{\textbf{h}}}_k^H + \Delta {{\textbf{h}}_k}\Delta {{\textbf{h}}}_k^H$ represents the uncertainty in the matrix ${\hat {\textbf{H}}_k}$.}

{ It is noted that ${\Delta _k}$ is a norm-bounded matrix as $\left\| {{\Delta _k}} \right\|_F \le {\xi _k}$. We can straightforwardly find the following relation \cite{SDR} as
\begin{equation*} \label{p22}
\begin{split}
\displaystyle  \left\| {{\Delta _k}} \right\|_F &  = \| {{{\hat {{\textbf{h}}}}_k}\Delta {{\textbf{h}}}_k^H + \Delta {{\textbf{h}}_k}\hat {{\textbf{h}}}_k^H + \Delta {{\textbf{h}}_k}\Delta {{\textbf{h}}}_k^H} \|_F\\
  \displaystyle & \le \| {{{\hat {{\textbf{h}}}}_k}\Delta {{\textbf{h}}}_k^H} \|_F{\rm{ + }}\| {\Delta {{\textbf{h}}_k}\hat {{\textbf{h}}}_k^H} \|_F{\rm{ + }}\| {\Delta {{\textbf{h}}_k}\Delta {{\textbf{h}}}_k^H} \|_F\\
 \displaystyle & \le  {\|{\hat {{\textbf{h}}}}_k \|}   \| {\Delta {{\textbf{h}}}_k^H} \| + \|\Delta {{\textbf{h}}_k}\| \| {\hat {{\textbf{h}}}_k^H} \|{\rm{ + }}{\| {\Delta {{\textbf{h}}_k}} \|^2}\\
 \displaystyle  &  =  \varepsilon _k^2 + 2{\varepsilon _k}\|{{\hat {{\textbf{h}}}}_k}\|,
\end{split}
\end{equation*}
where the first inequality is based on the triangle inequality, and
the second inequality come from the Cauchy-Schwarz inequality. It is possible to choose ${\xi _k} \triangleq \varepsilon _k^2 + 2{\varepsilon _k}\|{\hat {{\textbf{h}}}_k}\|$. It is noted that the bounds of this uncertainty are derived by triangle inequality, Cauchy-Schwarz inequality, and multiplicity of the second norm,  which are tight enough.}

Adopting the preceding notations, we can rewrite \eqref{p20} at the $n$-th iteration  as
\begin{equation} \label{p23}
\begin{split}
& \displaystyle \mathop {\min }\limits_{{\{{{\textbf{{V}}}}_{k}\}, { {\kern 1pt} {{\rho _k}} }, {\kern 1pt} \tau, {\kern 1pt}x_k, {\kern 1pt}{y_{{k}}}}}    {\kern 17pt} \tau \\
& \displaystyle {\rm{s.t.}}  ~   \mathop {\min }\limits_{\left\| {{\Delta _k}} \right\|_F \leq {\xi _k}} {\rm{tr}}\big( {( {{{\hat {{\textbf{H}}}}_k} + {\Delta _k}} ){{\textbf{{V}}}_k}} \big) \geq e^{x_k}, \\
&\!\!\!\!\!\! \displaystyle   \mathop {\max }\limits_{\left\| {{\Delta _k}} \right\|_F \leq {\xi _k}}  \sum\limits_{j \ne k} {\rm{tr}} ( {( {{{\hat {{\textbf{H}}}}_k} {\rm{+}} {\Delta _k}} ){{\textbf{{V}}}_j}} \big)  {\rm{+}} \sigma _k^2 {\rm{+}} \frac{\delta _k^2}{{\rho _k}} \leq  e^{{y}_{{k}}^{(n)}}({{y}_{{k}}} {\rm{-}} {{y}_{{k}}^{(n)}}{\rm{+}}1), \\
&\displaystyle  \mathop {\min }\limits_{\left\| {{\Delta _k}} \right\|_F \leq {\xi _k}} \sum\nolimits_{j = 1}^K {\rm{tr}}( {( {{{\hat {{\textbf{H}}}}_k} + {\Delta _k}} ){{\textbf{{V}}}_j}} \big)  \ge \frac{{e_k}}{{\zeta _k}\left( {1 - {\rho _k}} \right)} - \sigma _k^2, \\
& \eqref{p8c}, ~\eqref{p8d},~ \eqref{p20a}, ~{{\textbf{{V}}}_k} \succeq \textbf{0},~ {\rm{rank}}({{\textbf{{V}}}_k}) = 1.
\end{split}
\end{equation}
%% shuanglan
%\begin{equation} \label{p23}
%\begin{split}
%& \displaystyle \mathop {\max }\limits_{{\{{{\textbf{{V}}}}_{k}\}, { {\kern 1pt} {{\rho _k}} }, {\kern 1pt} \tau, {\kern 1pt}x_k, {\kern 1pt}{y_{{k}}}}}    {\kern 17pt} \tau \\
%& \displaystyle {\rm{s.t.}}      \mathop {\min }\limits_{\left\| {{\Delta _k}} \right\| \leq {\xi _k}} {\rm{tr}}\big( {( {{{\hat {{\textbf{H}}}}_k} + {\Delta _k}} ){{\textbf{{V}}}_k}} \big) \geq e^{x_k}, \\
%& \displaystyle   \mathop {\max }\limits_{\left\| {{\Delta _k}} \right\| \leq {\xi _k}}  \sum\limits_{j \ne k} {\rm{tr}} ( {( {{{\hat {{\textbf{H}}}}_k} {\rm{+}} {\Delta _k}} ){{\textbf{{V}}}_j}} \big)  {\rm{+}} \sigma _k^2 {\rm{+}} \frac{\delta _k^2}{{\rho _k}} \leq  e^{\hat{y}_{{k}}}({{y}_{{k}}} {\rm{-}} {\hat{y}_{{k}}}{\rm{+}}1), \\
%& \displaystyle  \mathop {\min }\limits_{\left\| {{\Delta _k}} \right\| \leq {\xi _k}} \sum\nolimits_{j = 1}^K {\rm{tr}}( {( {{{\hat {{\textbf{H}}}}_k} + {\Delta _k}} ){{\textbf{{V}}}_j}} \big)  \ge \frac{{e_k}}{{\zeta _k}\left( {1 - {\rho _k}} \right)} - \sigma _k^2, \\
%& ~\eqref{p8c}, ~\eqref{p8d},~ \eqref{p19_2}, ~{{\textbf{{V}}}_k} \succeq \textbf{0},~ {\rm{rank}}({{\textbf{{V}}}_k}) = 1.
%\end{split}
%\end{equation}
{Note that problem \eqref{p23} is non-convex due to the existence of ${\rm{tr}}( {( {{{\hat {{\textbf{H}}}}_k} + {\Delta _k}} ){{\textbf{{V}}}_k}} )$ in
both the SINR and EH constraints.} %It can be observed  that the mathematical expressions with ${\rm{tr}}( {( {{{\hat {{\textbf{H}}}}_k} + {\Delta _k}} ){{\textbf{{V}}}_k}} )$ exist in problem \eqref{p23}.}
 For computing ${\rm{tr}}( {( {{{\hat {{\textbf{H}}}}_k} + {\Delta _k}} ){{\textbf{{V}}}_k}} )$,  we have the following proposition.

\emph{\underline{\textbf{Proposition 1}} :} Let us denote $\Delta _k^{\min }$ and $\Delta _k^{\max } $ as the minimizer and the maximizer of ${\rm{tr}}( {( {{{\hat {{\textbf{H}}}}_k} + {\Delta _k}} ){{\textbf{{V}}}_k}} )$, respectively. Then, $\Delta _k^{\min }$ and $\Delta _k^{\max } $ are expressed as
\begin{equation} \label{p24}
 \Delta _k^{\min } =  - {\xi _k}\frac{{{\textbf{{V}}}_k^H}}{{\left\| {{\textbf{{V}}}_k} \right\|_F}},
~ \Delta _k^{\max } = {\xi _k}\frac{{{\textbf{{V}}}_k^H}}{{\left\| {{\textbf{{V}}}_k} \right\|_F}}.
\end{equation}

\emph{{\underline{Proof}:}} See Appendix B.  $ {\kern 130pt} \blacksquare$

{Using these results in  \eqref{p24} to remove the channel uncertainty ${\Delta _k}$, we get the following convex form as}
\begin{equation*} \label{p31}
\begin{split}
&\mathop {\min }\limits_{\left\| {{\Delta _k}} \right\|_F \leq {\xi _k}} \sum\limits_{j = 1}^K {\rm{tr}}\big( {( {{{\hat {{\textbf{H}}}}_k} {\rm{+}} {\Delta _k}} ){{\textbf{{V}}}_j}} \big)
=  \sum\limits_{j=1}^K {\big( {{\rm{tr}}( {{{{\bf{\hat {{\textbf{H}}}}}}_k}{{\textbf{{V}}}_j}} ){\rm{ - }}{\xi _k}\left\| {{{\textbf{{V}}}_j}} \right\|_F} \big)},\\
&\mathop {\max }\limits_{\left\| {{\Delta _k}} \right\|_F \leq {\xi _k}} \sum\limits_{j \ne k} {{\rm{tr}}\big( {( {{{{\bf{\hat {{\textbf{H}}}}}}_k} {\rm{+}} {\Delta _k}} ){{\textbf{{V}}}_j}} \big)}
{\rm{ = }}\sum\limits_{j \ne k} {\big( {{\rm{tr}}( {{{{\bf{\hat {{\textbf{H}}}}}}_k}{{\textbf{{V}}}_j}} ){\rm{ + }}{\xi _k}\left\| {{{\textbf{{V}}}_j}} \right\|_F} \big)}.
\end{split}
\end{equation*}
Thus, by removing the rank-one constraint, the associated SINR maximization problem can be rewritten as
\begin{subequations}\label{p32}
\begin{eqnarray}
&&\!\!\!\!\! \mathop {\min }\limits_{{\{{{\textbf{{V}}}}_{k}\}, { {\kern 1pt} {{\rho _k}} }, {\kern 1pt} \tau, {\kern 1pt}x_k, {\kern 1pt}{y_{{k}}}}}    {\kern 17pt} \tau \nonumber \\
&&\!\!\!\!\!  {\rm{s.t.}} ~~ {\rm{tr}}({{{\bf{\hat {\textbf{H}}}}}_k}{{\textbf{{V}}}_k}) - {\xi _k}\left\| { {{{\textbf{{V}}}_k}} } \right\|_F \ge   {e^{{x_k}}}, \label{p32a}\\
&&\!\!\!\!\!\!\!\!\!\!\!\!\!\!\!\!\!\!\!\!\!\!\!\!\!\!\!\!\!\!\!  \sum\limits_{j \ne k} {\big( {{\rm{tr}}( {{{{\bf{\hat {{\textbf{H}}}}}}_k}{{\textbf{{V}}}_j}} ){\rm{ + }}{\xi _k}\left\| {{{\textbf{{V}}}_j}} \right\|_F} \big)}  {\rm{+}} \sigma _k^2 {\rm{+}} \frac{\delta _k^2}{{\rho _k}} \leq  e^{{y}_{{k}}^{(n)}}({{y}_{{k}}} {\rm{-}} {{y}_{{k}}^{(n)}}{\rm{+}}1),  \label{p32b} \\
&&\!\!\!\!\!\!  \sum\limits_{j=1}^K {\big( {{\rm{tr}}( {{{{\bf{\hat {{\textbf{H}}}}}}_k}{{\textbf{{V}}}_j}} ){\rm{ - }}{\xi _k}\left\| {{{\textbf{{V}}}_j}} \right\|_F} \big)}  \ge \frac{{e_k}}{{\zeta _k}\left( {1 - {\rho _k}} \right)} - \sigma _k^2, \label{p32c} \\
&&  \eqref{p8c},~\eqref{p8d}, ~ \eqref{p20a}, ~\textbf{V}_k \succeq \textbf{0}, \forall k. \nonumber
\end{eqnarray}
\end{subequations}

Problem \eqref{p32} becomes a convex form  for a given $\{{y}_{{k}}^{(n)}\} $, which can be solved by using CVX \cite{CVX}. In the SCA approach, the approximation with the current optimal solution can be updated iteratively until the constraint \eqref{p32b} hold with equality. The SCA algorithm is outlined in Algorithm 2 below. In Algorithm 2, the optimal solution to problem \eqref{p32} at the $n$-th iteration is defined as $\{{{\textbf{{V}}}}_k^{*(n)}\}$, which achieves a stable point when the SCA algorithm converges. %Now, we consider the tightness analysis for
%the problem (26) due to rank relaxation.

%Thus, provided that (25) is feasible for positive secrecy rates, there exists an optimal solution of rank-one, the proof is similar to
%that of Theorem 1.

%%% yuanabn 171001
\begin{algorithm}
\caption{Robust Iterative Algorithm Based on SCA}
\label{alg: SCA }
\begin{algorithmic}
\STATE {Initialize $\{{y}_{{k}}^{(n)}\}$ and set $n= 0$.}

\STATE \textbf{Repeat}

${\kern 10pt} {\kern 5pt}$ Solve problem \eqref{p32} with $\{{y}_{{k}}^{(n)}\}$ to obtain ${{\textbf{{V}}}}_k^{*(n)}$ and  $\tau^{*(n)}$ for $k = 1,...,K$.

${\kern 10pt}{\kern 5pt}$ Set ${{y}_k}^{(n+1)} = {y}_k^{(n)}$ for $k = 1,...K$.

${\kern 10pt} {\kern 5pt}$ Update $n\leftarrow n+1$.

\textbf{Until}  Convergence

If   rank$({{\textbf{{V}}}}_k^{*(n)}) = 1$,

${\kern 10pt} {\kern 5pt}$  compute $\{{\textbf{v}}_{k}^{*}\}$ by EVD of ${{\textbf{{V}}}}_k^*(n)$.

else

${\kern 10pt} {\kern 5pt}$  use the GR technique to find $\{{\textbf{v}}_{k}^{*}\}$ for $k = 1,...,K$.
\end{algorithmic}
\end{algorithm}

%% 新版本  171010
%\begin{algorithm}
%\caption{{Robust Iterative Algorithm Based SCA}}
%\label{alg: SCA }
%\begin{algorithmic}
%\STATE {{Initialize the iteration number $n= 0$, $\{\hat{y}_{{k}}(n)\}$ such that problem \eqref{p32} is feasible, and  a prescribed accuracy
%tolerance $\mu >0$.}}
%
%\textbf{{Iteration loop begin:}}
%
% \STATE ${\kern 10pt}$ {\textbf{1).} Solve problem \eqref{p32} with $\{\hat{y}_{{k}}(n)\}$ to obtain ${{\textbf{V}}}_k^*(n)$ and the optimal objective value $\tau^*(n)$ for $k = 1,...,K$.}
%
% \STATE ${\kern 10pt}$ {\textbf{2).} Set ${\hat{y}_k}(n+1) = {\hat{y}_k}(n)$$, k = 1,...K$ and update the iteration number $n = n+1$.}
%
%
%{\textbf{Iteration loop end Until}}   {$|\tau^*(n+1) - \tau^*(n)| \leq \mu$.}
%
%{If rank$({{\textbf{V}}}_k^*(n)) = 1$, compute $\{{\textbf{v}}_{k}^{*}\}$ by EVD of ${{\textbf{V}}}_k^*(n)$. If rank$({{\textbf{V}}}_k^*(n)) = 2$,  use Gaussian randomization technique to find $\{{\textbf{v}}_{k}^{*}\}$ for $k = 1,...,K$ \cite{SDR}.}
%
%\end{algorithmic}
%\end{algorithm}
\section{Computational Complexity}
In this section, we evaluate the computational complexity of the proposed robust design methods. As will be shown in Section V, the proposed algorithms exhibit gains in terms of both computational complexity and performance compared to the conventional SDP scheme in \cite{zhu_15VTC_SWIPT_DAS} which employs local search. Now, we will present the complexity comparison by adopting the analysis in \cite{Complexity} and \cite{Complexity_zhu}. The complexities of the proposed algorithms are shown in Table I.
Here,  we denote $n$, $L^{{\rm{max}}} =
{{{\log }_2} {\frac{{{\Gamma_{\max }} }}{\eta }} } $, $Q^{{\rm{max}}}$ and $D^{{\rm{max}}}$   as the number of decision variables, the bisection search number, the SCA iteration number and the local search number in \cite{zhu_15VTC_SWIPT_DAS}, respectively.

\emph{1)  Algorithm 1} in problem \eqref{p17_5} involves $2K$ LMI constraints of size $MN_{{T}}+2$, $K$ LMI constraints of size $MN_{{T}}$,  and  $4K+M$ linear constraints. %% $M+K$ times eigenvalue decomposition of

\emph{2)  Algorithm 2} in problem \eqref{p32} has $K$ second-order
cones (SOC) constraints of dimension $M^2N_{{T}}^2+1$, $K$ SOC constraints of dimension $(K-1)M^2N_{{T}}^2+1$,  $K$ SOC constraints of dimension $KM^2N_{{T}}^2+1$, $K$ LMI constraints of size $MN_{{T}}$, and $3K+M$ linear constraints.

\emph{3) Conventional  scheme} in \cite{zhu_15VTC_SWIPT_DAS} consists of $2K$ LMI constraints of size $MN_{{T}}+1$, $K$ LMI constraints of size $MN_{{T}}$, and $2K+M$ linear constraints.

For example, for a system with $M = 3, K = 2, N_T = 3$,  $L^{{\rm{max}}} = Q^{{\rm{max}}} = 6$, and $D^{{\rm{max}}} = 100$, the complexities of the proposed Algorithm 1, Algorithm 2, and the conventional scheme \cite{zhu_15VTC_SWIPT_DAS} are ${\cal O}(1.96\times 10^9)$, ${\cal O}(3.41\times 10^8)$ and ${\cal O}(4.31 \times 10^{10})$, respectively. Thus the complexity of the proposed Algorihm 1 and Algorihm 2  are only $4.5\%$ and $0.8\%$  of that of the conventional scheme in \cite{zhu_15VTC_SWIPT_DAS}, respectively  .

\begin{table*}[htbp]
\caption{ Complexity analysis of different algorithms}
\label{tab:threesome}
\centering
\begin{tabular}{|c|c|}
\hline
 Algorithms & Complexity Order   \\
\hline \hline
 Algorithm 1 & $\begin{array}{l}  {\cal O}\big( nL^{{\rm{max}}}Q^{{\rm{max}}} \sqrt {2K(MN_T{\rm{+}}2) {\rm{+}} KMN_T {\rm{+}} 4K{\rm{+}}M}  \big\{2K(MN_T {\rm{+}} 2)^3{\rm{+}} K(MN_T)^3 {\rm{+}}n[2K(MN_T \\ {\rm{+}} 2)^2{\rm{+}}K(MN_T)^2]{\rm{+}}4K{\rm{+}}M{\rm{+}}n^2  \big\}\big) {\textrm{where}} ~ n ={\cal O}({M^2N^2_T} {\rm{+}} 3K {\rm{+}} 1) \end{array} $ \\
\hline
Algorithm 2 & $ \begin{array}{l} {\cal O}\big( n{Q^{max}}\sqrt {6K {\rm{+}} KMN_T{\rm{+}}3K{\rm{+}}M} \big\{K[(M^2N_T^2  {\rm{+}}1)^2{\rm{+}}((K{\rm{-}}1)M^2N_T^2{\rm{+}}1)^2{\rm{+}}(KM^2N_T^2{\rm{+}}1)^2] \\ {\rm{+}}K[(MN_T)^3{\rm{+}}n(MN_T)^2]{\rm{+}}3K{\rm{+}}M{\rm{+}}n^2\big\}\big) {\textrm{where}} ~ n ={\cal O}(M^2N^2_T {\rm{+}} 3K {\rm{+}} 1) \end{array} $\\
\hline
Conventional  scheme \cite{zhu_15VTC_SWIPT_DAS} & $ \begin{array}{l} {\cal O}\big( n{D^{max}}\sqrt {K(3MN_{{T}}{\rm{+}}2) {\rm{+}} 2K{\rm{+}}M} [2K(MN_{{T}}  {\rm{+}} 1)^3 {\rm{+}} KM^3N^3_{{T}} {\rm{+}} n(2K(MN_{{T}}{\rm{+}}1)^2 {\rm{+}} KM^2N_{{T}}^2 \\ {\rm{+}}2K{\rm{+}}M) {\rm{+}}n^2]\big)  {\textrm{where}} ~ n ={\cal O}(M^2N^2_T {\rm{+}} 2K {\rm{+}} 1) \end{array} $\\
\hline
\end{tabular}
\end{table*}
\section{Simulation Results}
In this section, we numerically compare the performance of the proposed algorithms for multiuser DAS SWIPT systems.
Throughout the simulation, we consider DAS with a circular antenna layout and set $M = 3, K = 3$, and $N_T =4$. The power of each DA port is set to ${P_1} = \frac{P}{{6}}$, ${P_2} = \frac{P}{{3}}$, and ${P_3} = \frac{P}{{2}}$ as in \cite{zhu_15VTC_SWIPT_DAS}.
Three DA ports form an equilateral triangle while all MSs are uniformly distributed inside a disc with the cell radius $R =\sqrt{\frac{112}{3}}$  m centered at the centroid of the triangle. The $j$-th DA port is located at $( { {r\cos {\frac{{2\pi ( {j - 1} )}}{M}}}, {\kern 5pt} {r\sin {\frac{{2\pi ( {j - 1} )}}{M}} } } )$ for $j=1,...,M$ with $ r= \sqrt{\frac{3}{7}}R$ as in \cite{Lee_12TWC_DAS}. The pathloss exponent $\gamma$ is set to be 3. According to this setting, a received SNR loss of 23.5 dB is observed at cell edge users compared to cell center users.
All channel coefficients ${\bar{\textbf{{h}}}_{m,k} \in \mathbb{C}^{N_T \times 1}}$ are modelled as Rician fading. The channel vector $\bar{\textbf{{h}}}_{m,k}$  is given as $\bar{\textbf{{h}}}_{m,k} = \sqrt{\frac{K_R}{1+K_R}}\bar{\textbf{{h}}}_{m,k}^{LOS} + \sqrt{\frac{1}{1+K_R}}\bar{\textbf{{h}}}_{m,k}^{NLOS}$,  where $\bar{\textbf{{h}}}_{m,k}^{LOS}$ indicates the line-of-sight (LOS) component with
$\| \bar{\textbf{{h}}}_{m,k}^{LOS} \|^2 = d_{m,k}^{ - {\gamma  \mathord{\left/ {\vphantom {\alpha  2}} \right. \kern-\nulldelimiterspace} 2}}$, $\bar{\textbf{{h}}}_{m,k}^{NLOS}$ represents the Rayleigh fading component as $\bar{\textbf{{h}}}_{m,k}^{NLOS} \sim \mathcal{CN}(0,d_{m,k}^{ - {\gamma  \mathord{\left/ {\vphantom {\alpha  2}} \right. \kern-\nulldelimiterspace} 2}}\mathbf{I})$, and $K_R$ is the Rician factor equal to 3. For the LOS component, we apply the far-field uniform linear antenna array to model the channels in \cite{LOS_channel}.  For simplicity, it is assumed that all MSs have the same set of parameters, i.e., ${\zeta _k} = \zeta, \delta _k^2 = {\delta ^2}, \sigma _k^2 = {\sigma ^2}$, and $e_k = e$ for $k = 1,...,K$. In addition, we set ${\sigma ^2} = {\rm{-50}}$ dBm, ${\delta ^2} = {\rm{-30}}$ dBm, and $\zeta = 0.3$. Also, all the channel uncertainties are chosen to be the same as $\varepsilon_k = \varepsilon, \forall k$. In the simulation, the worst-case rate in all the ID users $\mathop {\min }\limits_{\forall j} \mathop {\min }\limits_{\Delta {{\textbf{h}}_j} \in {{\cal H}_j}} \log_2(1+{{\rm{SINR}}_j}) $ is plotted by taking an average over 1000 randomly generated channel realizations. % We compare the following designs: perfect CSI case, the proposed algorithms, and the conventional local search algorithm \cite{zhu_15VTC_SWIPT_DAS}.

\begin{figure}[!t]
\begin{center}
\includegraphics[width=3.5in]{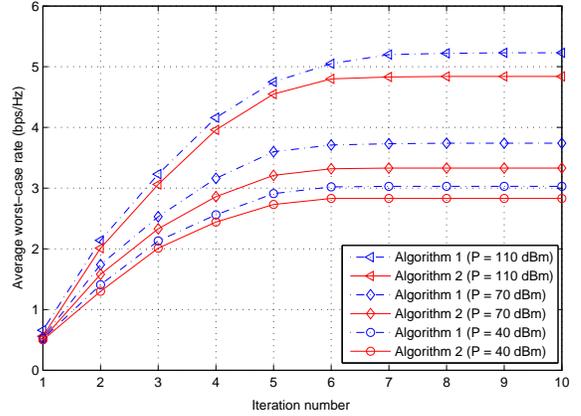}
\end{center}
\caption{Convergence performance of the proposed iterative algorithm for various $P$}
\end{figure}
Fig. 2 investigates the convergence performance of the proposed algorithms with $e$ = 3 dBm  and $\varepsilon=0.01$. %The initial PS factors $\rho_k$ for Algorithm 1 is chosen as $\rho_k=0.5, \forall k$. %In Fig. 3, we present the average worst-case SINR achieved by the proposed iterative optimization algorithm in terms of iterations with different numbers of DA ports.
It is clear that the proposed iterative algorithms indeed converge in all cases. We can see that after 7 iterations, the steady average worst-case rate is achieved for all $P$.

\begin{figure}[!t]
\begin{center}
\includegraphics[width=3.5in]{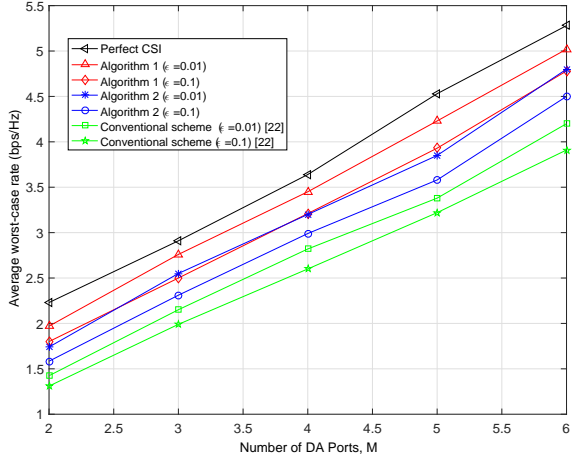}
\end{center}
\caption{Average worst-case rate versus the number of DA ports}
\end{figure}

In Fig. 3, we present the average worst-case rate versus the number of DA ports $M$ with various channel uncertainty $\varepsilon$ with $P$ = 60 dBm, $e$ = 5 dBm  and $\varepsilon=0.01$. It is found that our proposed robust algorithms attain substantial worst-case rate improvements over the conventional scheme in \cite{zhu_15VTC_SWIPT_DAS}. It is observed that there is about 0.3 bps/Hz difference between the curves of $\varepsilon$ = 0.01 and 0.1 for the proposed algorithms.  Furthermore, our proposed Algorithm 2 achieves about 0.5 bps/Hz and 0.7 bps/Hz gain compared to the conventional scheme \cite{zhu_15VTC_SWIPT_DAS} for $\varepsilon$ = 0.01 and 0.1, respectively. We also see that our proposed Algorithm 1 outperforms Algorithm 2 at the expense of increased complexity.

 Fig. 4 shows the performance comparison among robust algorithms for different number of antennas in each DA port with  $e = 5$ dBm and $P$  = 80 dBm. One can see that the conventional algorithm \cite{zhu_15VTC_SWIPT_DAS} requires more antennas than our proposed robust algorithms. The performance gap between our proposed Algorithm 1 and 2 curves is about 0.3 bps/Hz. Moreover, as $N_T$ increases,  the performance gap between our proposed algorithms and the conventional scheme becomes bigger.
%% average worst-case rate

\begin{figure}[!t]
\begin{center}
\includegraphics[width=3.5in]{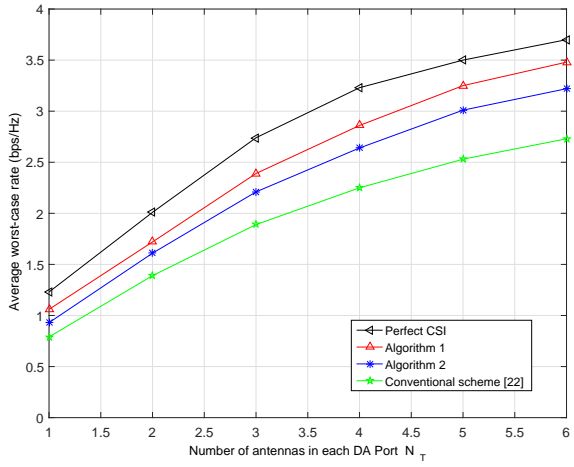}
\end{center}
\caption{Average worst-case rate versus the number of antennas in each DA port}
\end{figure}

\begin{figure}[!t]
\begin{center}
\includegraphics[width=3.5in]{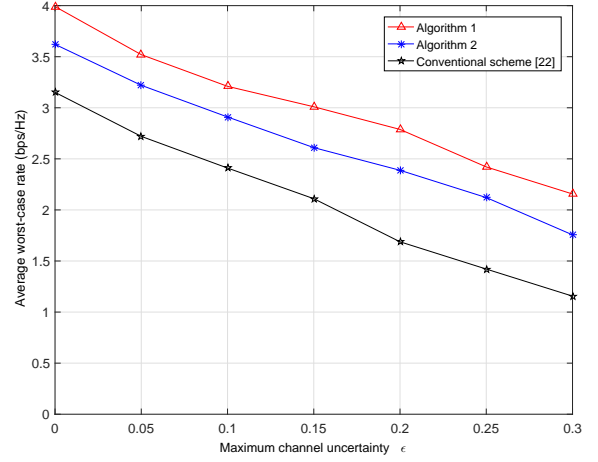}
\end{center}
\caption{Average worst-case rate versus channel uncertainty $\varepsilon$}
\end{figure}

Fig. 5 depicts the effect of the channel uncertainty $\varepsilon$ on the average worst-case rate with  $e = 0$ dBm and $P = 50$ dBm. We can check that as the maximum channel uncertainty $\varepsilon$ decreases, the average worst-case rate becomes enhanced. Clearly, the proposed robust algorithms outperform the conventional scheme \cite{zhu_15VTC_SWIPT_DAS}.  %%% OK.

\begin{figure}[!t]
\begin{center}
\includegraphics[width=3.5in]{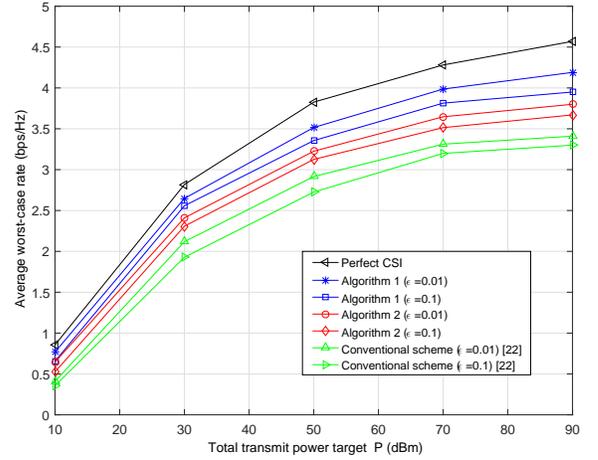}
\end{center}
\caption{Average worst-case rate versus $P$ for various $\varepsilon$}
\end{figure}

Finally, in Fig. 6, we exhibit the average worst-case rate versus the total transmit power target $P$ for various $\varepsilon$ with  $e = 3$ dBm.   Compared to our proposed Algorithm 1, Algorithm 2 achieves a complexity reduction at the expense of a performance loss. It is observed that as $\varepsilon$ increases, the performance gap between our proposed algorithms and the conventional scheme becomes larger.

\section{Conclusion}
In this paper, we have studied a design of robust transmit beamforming and power splitting in multiuser DAS SWIPT
downlink systems under per-DA port power constraint and energy harvesting constraint. Assuming imperfect CSIT, the uncertainty of the channel is modeled by an Euclidean ball. We have developed an algorithm to find a robust beamforming solution for maximizing the worst-case SINR by addressing a set of convex per-DA port power balancing problems. The reformulated problem can be solved by applying the SDR technique. %, which has been proved that the SDR is tight.
Also, given the beamforming solution, the PS factor has been calculated. We have proposed an iterative algorithm and a low-complexity algorithm for the worst-case SINR maximization problem.
%To reduce the computational complexity of local optimal solution, we propose two non-iterative algorithms to find suboptimal PS solution by using search algorithms over the max-min SINR values.
Simulation results have demonstrated the validity of the proposed algorithms.

%We have developed an efficient algorithm to find the optimal robust beamforming design for maximizing the worst-case SINR by addressing a set of convex
%per-DA port power balancing problems. The reformulated problem can be optimally solved by applying the SDR technique, which has been proved that the SDR is tight. The PS factor has been computed through the proposed search algorithms. % The suboptimal power splitting factor design with lower complexity is also presented, and its performance is compared against the proposed optimal solution.
%Simulation results have demonstrated the validity of the proposed methods.

%\section*{Acknowledgment}
%The work is supported by the National High Technology Research and Development Program ``863" of China (2014AA01A705), and the National Nature Science Foundation of China under grant 61172086, U1204607, 61201251; and Zhengyu Zhu also thank the support of China Scholarship Council (CSC).

%Using the local optimal beamforming solution, we obtain the optimal PS solution based on Lagrangian method and propose an iterative algorithm to find local optimum.
%To reduce the computational complexity of local optimal solution, we propose two non-iterative algorithms to find suboptimal PS solution by using search algorithms over the max-min SINR values.

\appendices
\section{Proof of Theorem 1}
If the rank-one constraint is ignored, problem \eqref{p17} becomes convex and satisfies the Slater's condition. Thus, its duality gap is zero \cite{convex}. Assume that the dual variables $\left\{ {{\rm{\textbf{C}}}{_k}} \right\} \in \mathbb{H}{_{\rm{ + }}},\left\{
{{\rm{\textbf{Q}}}{_k}} \right\} \in \mathbb{H}{_{\rm{ + }}},\left\{ {{\rm{\textbf{S}}}{_k}}
\right\} \in \mathbb{H}{_{\rm{ + }}}$
and $\left\{ {{\mu _m}} \right\} \ge
0$ correspond to the constraint
${{\rm{\textbf{A}}}_k}\succeq{\textbf{0}},{\kern 1pt} {\kern 1pt} {\kern 1pt}
{{\rm{\textbf{B}}}_k}\succeq{\textbf{0}},{{{\textbf{V}}}_k}\succeq{\textbf{0}}$ and $ {\sum\limits_{k = 1}^K {{\rm{tr}}({{{\textbf{D}}}_m}{{{\textbf{V}}}_k})}  \leq \alpha {P_m}} $ in \eqref{p17}, respectively.
%If the rank-one constraint is ignored, problem (14) becomes convex and satisfies the Slater's condition. Thus, its duality gap is zero [20]. Assume that the dual variables
%$\left\{ {{\rm{\textbf{C}}}{_k}} \right\} \in \mathbb{H}{_{\rm{ + }}},\left\{
%{{\rm{\textbf{Q}}}{_k}} \right\} \in \mathbb{H}{_{\rm{ + }}},\left\{ {{\rm{\textbf{S}}}{_k}}
%\right\} \in \mathbb{H}{_{\rm{ + }}}$ and $\left\{ {{\mu _m}} \right\} \ge
%0$ correspond to the constraint
%${{\rm{\textbf{A}}}_k}\succeq{\textbf{0}},{\kern 1pt} {\kern 1pt} {\kern 1pt}
%{{\rm{\textbf{B}}}_k}\succeq{\textbf{0}},{{{\textbf{V}}}_k}\succeq{\textbf{0}}$ and $ {\sum\limits_{k = 1}^K {{\rm{tr}}({{{\textbf{D}}}_m}{{{\textbf{V}}}_k})}  \leq \alpha {P_m}} $ in (14), respectively.
Then, the Lagrangian dual function of the primal problem \eqref{p17} is given by
%% danlan
%\begin{equation} \label{p55}
%{\cal L} =\alpha  - \sum\limits_{k = 1}^K {\left( {{\rm{tr}}\left( {{\rm{\textbf{C}}}{_k}{\kern 1pt} {{\rm{\textbf{A}}}_k}} \right) + {\rm{tr}}\left( {{\rm{\textbf{Q}}}{_k}{{\rm{\textbf{B}}}_k}} \right) + {\rm{tr}}\left( {{\rm{\textbf{S}}}{_k}{\kern 1pt} {{{\textbf{V}}}_k}} \right)} \right)}
% + \sum\limits_{m = 1}^M {{\mu _m}\left( {\sum\limits_{k = 1}^K {{\rm{tr}}({{{\textbf{D}}}_m}{{{\textbf{V}}}_k})}  - \alpha {P_m}} \right)}.
%\end{equation}
%%%  shuanglan
\begin{equation} \label{p55}
\begin{split}
{\cal L} &=\alpha  - \sum\limits_{k = 1}^K {\left( {{\rm{tr}}\left( {{\rm{\textbf{C}}}{_k}{\kern 1pt} {{\rm{\textbf{A}}}_k}} \right) + {\rm{tr}}\left( {{\rm{\textbf{Q}}}{_k}{{\rm{\textbf{B}}}_k}} \right) + {\rm{tr}}\left( {{\rm{\textbf{S}}}{_k}{\kern 1pt} {{{\textbf{V}}}_k}} \right)} \right)} \\
& + \sum\limits_{m = 1}^M {{\mu _m}\left( {\sum\limits_{k = 1}^K {{\rm{tr}}({{{\textbf{D}}}_m}{{{\textbf{V}}}_k})}  - \alpha {P_m}} \right)}.
\end{split}
\end{equation}

Since ${\rm{\textbf{C}}}{_k}$ and ${\rm{\textbf{T}}}{_k}$ are
Hermitian matrices, we have
\begin{equation*} \label{p56}
\begin{split}
& \displaystyle {\rm{tr}}({\rm{\textbf{C}}}{_k}{\kern 1pt} {{\rm{\textbf{A}}}_k}){\rm{ = tr}}({\rm{\textbf{C}}}{_k}{\kern 1pt} {\rm{\textbf{G}}}_k^H{\textbf{T}}{{\rm{\textbf{G}}}_k}) + {\rm{tr}}({\rm{\textbf{C}}}{_k}{{\rm{\textbf{F}}}_k}),  \\
& \displaystyle {\rm{tr}}({\rm{\textbf{Q}}}{_k}{\kern 1pt} {{\rm{\textbf{B}}}_k}){\rm{ =
tr}}({\rm{\textbf{Q}}}{_k}{\kern 1pt}
{\rm{\textbf{G}}}_k^H{{\textbf{M}}_k}{{\rm{\textbf{G}}}_k}) +
{\rm{tr}}({\rm{\textbf{Q}}}{_k}{{\rm{\textbf{E}}}_k}),  %\eqno{(33b)
\end{split}
\end{equation*}
where
\begin{equation*} \label{p56}
\begin{split}
& {{\rm{\textbf{E}}}_k} = \left[ \begin{array}{ccc}
\rho_{k} & \delta_k & \textbf{0}_{1 \times MN_{T}} \\
\delta_k &  \sigma_{k}^{2} {\rm{-}} r_{k} & {\textbf{0}_{1 \times MN_{T}}} \\
\textbf{0}_{MN_{T} \times 1} & {\textbf{0}_{MN_{T} \times 1}} & \frac{r_{k}}{\varepsilon_{k}^{2}}{\textbf{I}}
\end{array}
\right],\\
& {{\rm{\textbf{F}}}_k} = \left[ \begin{array}{ccc}
{\zeta _k}(1{\rm{-}}\rho_{k}) & \sqrt{{e}_{k}} & \textbf{0}_{1 \times MN_{T}} \\
\sqrt{{e}_{k}} & \sigma_{k}^{2} {\rm{-}} t_{k} & {\textbf{0}_{1 \times MN_{T}}} \\
\textbf{0}_{MN_{T} \times 1} & {\textbf{0}_{MN_{T} \times 1}} &  \frac{t_{k}}{\varepsilon_{k}^{2}}{\textbf{I}}
\end{array}\right], \\
& {{\rm{\textbf{G}}}_k} = [\begin{array}{*{20}{c}}
\textbf{0} & {{{\hat {{\textbf{h}}}}_k}} & {{\textbf{I}}}
\end{array}{\rm{]}}.
\end{split}
\end{equation*}

Taking partial derivative of  \eqref{p55} with respect to ${{{\textbf{V}}}_k}$ and applying the KKT conditions \cite{convex}, it follows
\begin{equation} \label{p57}
\sum\limits_{m = 1}^M {{\mu _m}{{{\textbf{D}}}_m}}- \left( {{\rm{\textbf{G}}}_k{\rm{\textbf{C}}}{_k}{\kern 1pt} {{\rm{\textbf{G}}}_k^H}{\rm{ + }}{\kern 1pt} {\textstyle{1 \over {\Gamma}}}{\rm{\textbf{G}}}_k{\rm{\textbf{Q}}}{_k}{{\rm{\textbf{G}}}_k^H}{\rm{ + \textbf{S}}}{_k}} \right){\kern 1pt}  = 0.
\end{equation}
Let $\left\{ {{\rm{\textbf{C}}}_k^ * } \right\},\left\{ {{\rm{\textbf{Q}}}_k^ * }
\right\},\left\{ {{\rm{\textbf{S}}}_k^ * } \right\}$ and $\left\{ {\mu _m^* } \right\}$ be the optimal dual solution to problem \eqref{p17}. Note that ${\rm{\textbf{Q}}}_k^ * {\rm{\textbf{B}}}_k^ *  = {\textbf{0}}$ from the complementary slackness conditions of problem \eqref{p17}. Since the size of ${\rm{\textbf{Q}}}_k^ * $ and ${\rm{\textbf{B}}}_k^ * $ is $\left( {MN_T + 2} \right) \times \left( {MN_T + 2} \right)$, we have ${\rm{rank}}({\rm{\textbf{Q}}}_k^ * ) + {\rm{rank}}({\rm{\textbf{B}}}_k^ * ) \le {MN_T + 2}$. Denoting $r _k^ *$ as the optimal solution to problem \eqref{p17}, $r _k^ *$ in ${\rm{\textbf{B}}}_k^ * $ in \eqref{p17} is non-negative. If $r _k^ *  > 0$, $r _k^ * {{\textbf{I}}}{\rm{ + }}{{{\textbf{M}_k^ *}} }$ has full rank. We will prove that $r _k^ *  \ne 0$ by contradiction.

If $r _k^ * =0$, the constraint $\|\Delta
{\textbf{h}}_k \|^2 \le \varepsilon _k^2$ does not hold since $r _k^ * $ is the dual variable for \eqref{p16}. Note that the
condition $\|\Delta
{\textbf{h}}_k \|^2 \le \varepsilon _k^2$ is the only constraint on $\Delta {\textbf{h}}_k$. If $\Delta {\textbf{h}}_k $ is the worst channel uncertainty which minimizes
$q \triangleq {{{{ \rho }_k}{{| {{\textbf{h}}_k^H{{\textbf{v}}_k}} |}^2}} \mathord{\left/
 {\vphantom {{{{ \rho }_k}{{\left| {{\textbf{h}}_k^H{{\textbf{v}}_k}} \right|}^2}} {\big( {{{ \rho }_k}\sum\nolimits_{j \ne k} {{{| {{\textbf{h}}_k^H{\textbf{{v}}_j}} |}^2}}  + {{ \rho }_k}\sigma _k^2 + \delta _k^2} \big)}}} \right.
 \kern-\nulldelimiterspace} {\left( {{{ \rho }_k}\sum\nolimits_{j \ne k} {{{| {\textbf{h}_k^H{\textbf{{v}}_j}} |}^2}}  + {{ \rho }_k}\sigma _k^2 + \delta _k^2} \right)}}$,
we can always find a scalar $\omega  > 1$ which satisfies $\|\Delta
{\textbf{h}}_k \|^2 =\varepsilon _k^2$.
Substituting the channel uncertainty $\omega \Delta {\textbf{h}}_k$ in
 $q$, we can find a SINR lower than that obtained by $\Delta {\textbf{h}}_k^*$. This is contradictory to the assumption that $\Delta {\textbf{h}}_k^*$ minimizes the SINR. Thus, it follows $r _k^ *  \ne 0$, which leads to $r _k^ *  > 0$.
As a result, $r _k^ * {\textbf{I}}{\rm{ +
}}{{\textbf{M}_k^ *} }$ becomes full rank, and we have ${\rm{rank}}(\textbf{B}_k^
* ) \ge N$. Furthermore, since ${\rm{rank}}({\rm{\textbf{Q}}}_k^ * )$ is non-zero. Thus, the rank of ${\textbf{Q}}_k^*$ equals 1. Similarly, we can show that ${\rm{rank}}\left( {{\rm{\textbf{C}}}_k^ * } \right)=1$. Then, it follows
${\rm{rank}}\big( {{\textbf{G}}_k^H( {{\textbf{C}}_k^ * {\kern 1pt}  + \frac{1}{{\Gamma}}{\textbf{Q}}_k^ * } ){{\textbf{G}}_k}} \big) \le {\rm{rank}}\big( {{\textbf{G}}_k^H{\textbf{C}}_k^* {\kern 1pt} {{\textbf{G}}_k}} \big)+ \frac{1}{{\Gamma}}{\rm{rank}}\big( {{\textbf{G}}_k^H{\textbf{Q}}_k^*{{\textbf{G}}_k}} \big) = 2$.

Thus, multiplying both sides of \eqref{p57} with ${\textbf{{V}}}_k^ * $ yields
\begin{equation*}
\big( {\sum\limits_{m = 1}^M {\mu _m^ * {{\textbf{D}}_m}} } \big){\textbf{{V}}}_k^ *
= \big( {{{\textbf{G}}}_k\left( {{{\textbf{C}}}_k^ * {\kern 1pt} + {\textstyle{1 \over {\Gamma}}}{\textbf{Q}}_k^* } \right){{{\textbf{G}}}_k^H}{{ + \textbf{S}}}_k^ * } \big){\textbf{{V}}}_k^ *,
\end{equation*}
where it is noted that ${\rm{\textbf{S}}}_k^ * {\textbf{{V}}}_k^ * =\textbf{0}$.
Since $\sum\limits_{m = 1}^M {{\mu _m^ *
{{\textbf{D}}_m}} } $ has full rank,
following the rank inequality ${\rm{rank}}(\textbf{A}\textbf{B})\leq \rm{min}({\rm{rank}}(\textbf{A}), {\rm{rank}}(\textbf{B}))$, we can finally prove that
\begin{equation*}
\begin{split}
\displaystyle
& {\rm{rank}}\big\{ {\big( {\sum\limits_{m = 1}^M {\mu _m^ * {{\textbf{D}}_m}} } \big){\textbf{{V}}}_k^*} \big\} = {\rm{rank}}\left( {{\textbf{{V}}}_k^ * } \right)\\
\leq & {\rm{rank}}\big( {{\textbf{G}}_k\big( {{\textbf{C}}_k^ * + \frac{1}{{\lambda}}{\textbf{Q}}_k^* } \big){{\textbf{G}}_k^H}} \big)
\leq 2.
\end{split}
\end{equation*}

\section{Proof of Proposition 1}
By introducing an arbitrary positive multiplier $\theta  \ge 0$, the Lagrangian function is given by
\begin{equation*} \label{p25}
 \!\!\!\!  {\rm{L}}\left( {{\Delta _k},\theta } \right) = {\rm{tr}}\big( {( {{{\hat {{\textbf{H}}}}_k} + {\Delta _k}} ){{\textbf{{V}}}_k}} \big) + \theta \big( {{{\left\| {{\Delta _k}} \right\|}^2} - \xi _k^2} \big).
\end{equation*}
We differentiate the Lagrangian function with respect to $\Delta _k^*$ and equate it to zero \cite{Matrix_analysis} as
\begin{equation*} \label{p26}
{\nabla _{\Delta _k^*}}{\rm{L}}\left( {{\Delta _k},\theta } \right){\rm{ = }}{\textbf{{V}}}_k^H + \theta {\Delta _k} = 0.
\end{equation*}
Then, we can find the optimal solution $\Delta _k^{{\rm{opt}}}{\rm{ = }} - \frac{1}{\theta }{\textbf{{V}}}_k^H.$
In order to remove the role of an arbitrary parameter of $\theta $, the Lagrangian function is differentiated with respect to $\theta $ and set to zero as
\begin{equation*} \label{p27}
{\nabla _\theta }{\rm{L}}\left( {{\Delta _k},\theta } \right){\rm{ = }}{\left\| {\Delta _k^{{\rm{opt}}}} \right\|^2} - \xi _k^2 = 0.
\end{equation*}
Thus the optimal solution for $\theta $ is obtained as
${\theta ^{{\rm{opt}}}}{\rm{ = }}\frac{{\| {{\textbf{{V}}}_k^H} \|}}{{{\xi _k}}}$.

By combining the above results, we finally get
\begin{equation*} \label{p28}
\Delta _k^{{\rm{opt}}}{\rm{ = }} \pm {\xi _k}\frac{{{\textbf{{V}}}_k^H}}{{\left\| {{{\textbf{{V}}}_k}} \right\|}}.
\end{equation*}
Accordingly, the minimum and maximum of $\Delta _k$ can be expressed as
\begin{equation*} \label{p29}
\Delta _k^{{\rm{min}}}{\rm{ = }} - {\xi _k}\frac{{{\textbf{{V}}}_k^H}}{{\left\| {{{\textbf{{V}}}_k}} \right\|}}, ~~
\Delta _k^{{\rm{max}}}{\rm{ = }}  {\xi _k}\frac{{{\textbf{{V}}}_k^H}}{{\left\| {{{\textbf{{V}}}_k}} \right\|}}.
\end{equation*}
To check if this optimal solution is a minimum, we confirm that the second derivative at the optimal solution point $\Delta _k^{{\rm{opt}}}$ is positive semi-definite as
\begin{equation*} \label{p30}
\nabla _{\Delta _k^*}^2{\rm{L}}\left( {\Delta _k^{{\rm{opt}}},{\theta ^{{\rm{opt}}}}} \right) = {\theta ^{{\rm{opt}}}}{\left( {{\rm{vec}}\left\{ {{{\bf{I}}_{MN_T}}} \right\}{\rm{vec}}\left\{ {{{\bf{I}}_{MN_T}}} \right\}} \right)^T}\succeq 0.
\end{equation*}

\ifCLASSOPTIONcaptionsoff
  \newpage
\fi

\end{document}